\documentclass[twoside]{raa} 
\usepackage{graphicx,times}
\usepackage{natbib}
\usepackage{amssymb,amsmath}
\usepackage{subfigure}
\usepackage{longtable}
\usepackage{verbatim}
\usepackage[section] {placeins}
\usepackage{rotating}
\bibpunct{(}{)}{;}{a}{}{,}
\usepackage[a4paper=true,dvipdfm=true,pagebackref=true]{hyperref}
\hypersetup{pdftitle = The title of my PDF, pdfauthor = My name, pdfsubject= The subject, pdfkeywords = keyword1 keyword2 keyword3}
\hypersetup{colorlinks = true, linkcolor = black, anchorcolor = red, citecolor = black, filecolor = red, pagecolor = red, urlcolor = red}

\begin{document}
   \title{Seismic diagnostics of solar-like oscillating stars}
 \volnopage{ {\bf 2012} Vol.\ {\bf X} No. {\bf XX}, 000--000}
   \setcounter{page}{1}
   \author{Yaguang Li\inst{1},  Minghao Du\inst{1}, Bohan Xie\inst{1}, Zhijia Tian\inst{2}, Shaolan Bi\inst{1}, Tanda Li\inst{3, 4}, Yaqian Wu\inst{1}, Kang Liu
      \inst{1}
   }
   \institute{ Department of Astronomy, Beijing Normal University, Beijing 100875,
China; {\it hnwilliam@hotmail.com; bisl@bnu.edu.cn}\\
\and
Department of Astronomy, Peking University, Beijing 100871, China; {\it tianzhijia@pku.edu.cn}\\
\and
Sydney Institute for Astronomy (SIfA), School of Physics, University of Sydney, NSW 2006, Australia;\\
\and
Stellar Astrophysics Centre, Department of Physics and Astronomy, Aarhus University, Ny Munkegade 120, DK-8000 Aarhus C, Denmark
\vs \no
}
\abstract{High precision and long-lasting \emph{Kepler} data enabled us to estimate stellar properties with asteroseismology as an accurate tool. We performed asteroseismic analysis on six solar-like stars observed by the \emph{Kepler} mission: KIC 6064910, KIC 6766513, KIC 7107778, KIC 10079226, KIC 10147635 and KIC 12069127. The extraction of seismic information includes two parts. First, we obtained two global asteroseismic parameters, mean large separation $\Delta\nu$ and frequency of maximum power $\nu_{\rm{max}}$, with autocorrelation function and collapsed autocorrelation function. Second, we extracted individual oscillation modes $\nu_{nl}$ with low-$l$ degree using a least-squares fit. Stellar grid models were built with Yale Rotating Stellar Evolution Code (YREC) to analyze stellar properties. They covered the range of $M=0.8\sim1.8$ $M_{\odot}$ with a step of 0.02 $M_{\odot}$ and $\rm{[Fe/H]}=-0.3\sim0.4$ $\rm{dex}$ with a step of 0.1 $\rm{dex}$. We used a Bayesian approach to estimate stellar fundamental parameters of the six stars, under the constraints of asteroseismic parameters ($\Delta\nu$, $\nu_{\rm{max}}$) and non-asteroseismic parameters ($T_{\rm{eff}}$, $\rm{[Fe/H]}$). We discover that the six targets include five sub-giant stars with $1.2\sim1.5$ $M_{\odot}$ and one main-sequence star with $1.08M_{\odot}$, and with ages in the range of $3\sim5$ $\rm{Gyr}$.
}
   \authorrunning{Y. G. Li et al. }            
   \titlerunning{Seismic diagnostics of solar-like oscillating stars}  
   \maketitle

\section{Introduction}
\label{sect:intro}

Recent space missions, such as CoRoT and \emph{Kepler} mission, have led us to a golden epoch when large scale asteroseismic analysis of stars can be carried out. Thanks to the high precision and long-lasting observation provided by these space missions, new previously unavailable areas of frequency domain have been opened (e.g. \citealt{2008A&A...488..705A}; \citealt{2007AAS...21011007B}; \citealt{2010arXiv1011.0435G}; \citealt{2013MNRAS.435..242G}). With detected oscillation, the following aseteroseismic studies are able to provide us a unique approach to constrain the star's fundamental properties, and even to test the theory of stellar structure and evolution. They enrich our knowledge not only on stars, but also on clusters and the Galaxy, or even broader, the whole universe (e.g.  \citealt{2007A&A...471..885S}; \citealt{2013ApJ...763...49D}; \citealt{2015EPJWC.10102004C}; \citealt{2015EPJWC.10101015C}; \citealt{2016ApJ...822...15S}).

Solar-like oscillations refer to stars oscillating with the same mechanism as the Sun, where they are stochastically excited and damped by convection motion in the near-surface convection zone (e.g. \citealt{1982AdSpR...2...11C}; \citealt{1983SoPh...82..469C}; \citealt{1999A&A...351..582H}). Study of oscillations could yield worthful conceptions on stellar structures and evolutionary stages. Main-sequence stars behave as p modes (pressure dominated) in the envelope. Sub-giant stars behave as mixed modes, which are characterized by g modes (gravity dominated) in the core and p modes in the envelope \citep{1980ApJS...43..469T}, when \emph{``avoided crossing''} commence (\citealt{1977A&A....58...41A}; \citealt{2014ApJ...781L..29B}; \citealt{2015A&A...580A.141L}). Therefore, oscillations are capable of distinguishing different types of stars with their identical signatures. Mixed modes have further shown potential on constraining stellar models in a powerful way (e.g. \citealt{2012ApJ...756...19D}; \citealt{2013EPJWC..4303002M}; \citealt{2013ApJ...769..141S}; \citealt{2014A&A...572L...5M}; \citealt{2015EPJWC.10101001M}), since some stellar parameters are particularly sensitive to them, e.g. stellar age (\citealt{2010ApJ...723.1583M}) and mass (\citealt{2012ApJ...745L..33B}). It is even possible to determine the presence and size of the convective core with the help of asteroseismology (\citealt{2014ApJ...780..152L}; \citealt{2015MNRAS.453.2094Y}).

Accurate data analysis on oscillation is one crucial prerequisite for detailed stellar diagnostics (e.g. \citealt{2010Ap&SS.328...67O}; \citealt{2012ApJ...756...19D}; \citealt{2013ApJ...769..141S}; \citealt{2014IAUS..298..375S}; \citealt{2014ApJS..210....1C}). Two global asteroseismic parameters, mean large separation $\Delta\nu$ and frequency of maximum power $\nu_{\rm{max}}$ which reflect the star's properties, have been designed to extract through pipelines (e.g. \citealt{2014ApJS..210....1C}; \citealt{2011MNRAS.415.3539V}; \citealt{2016MNRAS.456.2183D}), which is even possible when the signal to noise ratio (S/N) is not that high (e.g. \citealt{2008A&A...485..813C}, \citealt{2009ApJ...700.1589S}). Individual oscillation frequencies, which can provide more insights related to a star's interior, have been broadly analyzed when convolved with maximum likelihood estimators and Bayesian estimators (e.g. \citealt{2016MNRAS.456.2183D}; \citealt{2014A&A...566A..20A}; \citealt{2012A&A...543A..54A}). In this work, we aim to analyze a low-mass sample of main-sequence stars and sub-giant stars by deriving $\Delta\nu$, $\nu_{\rm{max}}$, and individual oscillation frequencies. This approach is expected to detect low angular degree (e.g. $l=0,1,2$) modes while modes with higher angular degree remain invisible due to geometrical cancellation. With global seismic parameters, deriving stellar properties would be an obvious and simple way to investigate these stars.

The article is organized as follows. In section \ref{sect:Obs}, we briefly introduce our selection of targets. In section \ref{sect:aa}, we illustrate the process of derivation of two seismic parameters $\Delta\nu$ and $\nu_{\rm{max}}$, and individual oscillation frequencies. In section \ref{sect:grid}, we present stellar model construction and their usage in estimating stellar fundamental parameters of the six stars. Finally, discussions and conclusions are shown in section \ref{sect:dac}.

\section{List of targets}
\label{sect:Obs}

We revisited the topic explored by \cite{2014ApJS..210....1C}, who derived the values of mean large separation $\Delta\nu$ and frequency of maximum power $\nu_{\rm{max}}$. We selected six targets which met the following criteria: S/N values are high enough to obtain individual oscillation frequencies, and they have not been extracted in any work before. We intend to obtain $\Delta\nu$ and $\nu_{\rm{max}}$ in our own way to test if our method works well, and then derive oscillation frequencies with which detailed asteroseismic diagnostics could be realized.

The \emph{Kepler} mission provides photometric time series of the targets with long cadence (LC; 29.43 min sampling) and short cadence (SC; 58.84s sampling). The pulsation frequency range is estimated to be above the Nyquist frequency of LC data. Here, we obtained SC time series over one year, which were collected from the \emph{Kepler} Asteroseismic Science Consortium website\footnote{http://kasoc.phys.au.dk/}. They had been preprocessed by the \emph{Kepler} Working Group (WG\#1, \citealt{2011MNRAS.414L...6G}). Table \ref{tab1} shows \emph{Kepler} mission data we used in this work.

Atmospheric parameters of the stars are crucial since they serve as constraints on stellar models. We noticed that the six targets were covered by LAMOST-\emph{Kepler} project, and were observed by LAMOST low resolution ($\sim$1800) optical spectra in the waveband of 3800$\sim$9000${\AA}$ by September 2014. Three atmospheric parameters, $T_{\rm{eff}}$, $\log{g}$, and $\rm{[Fe/H]}$ were derived through the LAMOST Stellar Parameter Pipeline (LSP3, \citealt{2015yCat..74480822X}).

Table \ref{tab1} also shows atmospheric parameters derived from \cite{2015yCat..74480822X}. Instead, LSP3 $\log{g}$ are found to exhibit non-negligible systematic bias. \cite{2016RAA....16...45R} claimed to have good agreement with asteroseismic results, but a 0.2 dex dispersion is quite large for the usage in this work. This is likely a consequence of the algorithm (weighted mean) and the template (MILES empirical library) that LSP3 adopted for the parameter estimation. The LSP3 [Fe/H] estimates are found to exhibit only minor systematic bias according to examinations with [Fe/H] from high resolution spectroscopy and with [Fe/H] of member stars of open clusters \citep{2015MNRAS.448...90X,2016MNRAS.tmp.1525X}. $T_{\rm{eff}}$ do not possess too much bias either \citep{2015MNRAS.454.2863H}. Therefore, we depleted $\log{g}$ from LSP3 in our analysis.

\section{Data analysis}
\label{sect:aa}
\subsection{Preprocessing of data}

First, for the six targets, we concatenated all the time series of the six targets and preprocessed them using the method described by \cite{2011MNRAS.414L...6G}, correcting outliers, jumps and drifts on the flux, and then passed the light curve through a high-pass filter with width of one day. The high-pass filter was built based on a moving-average smoothing function with Gaussian weights. It only affects frequencies lower than 11.57$\mu\rm{Hz}$, away from the oscillation frequency range we intend to analyze. We then normalized them by dividing by the mean value of each series. This will make each series indistinguishable.

Second, we obtained the power spectra of the six targets by applying Lomb-Scargle Periodogram (\citealt{1976Ap&SS..39..447L}; \citealt{1982ApJ...263..835S}) method, which is especially suitable for irregular spaced discrete data with gaps.

\begin{figure}
\centering
     \includegraphics[width=7.0cm, angle=0]{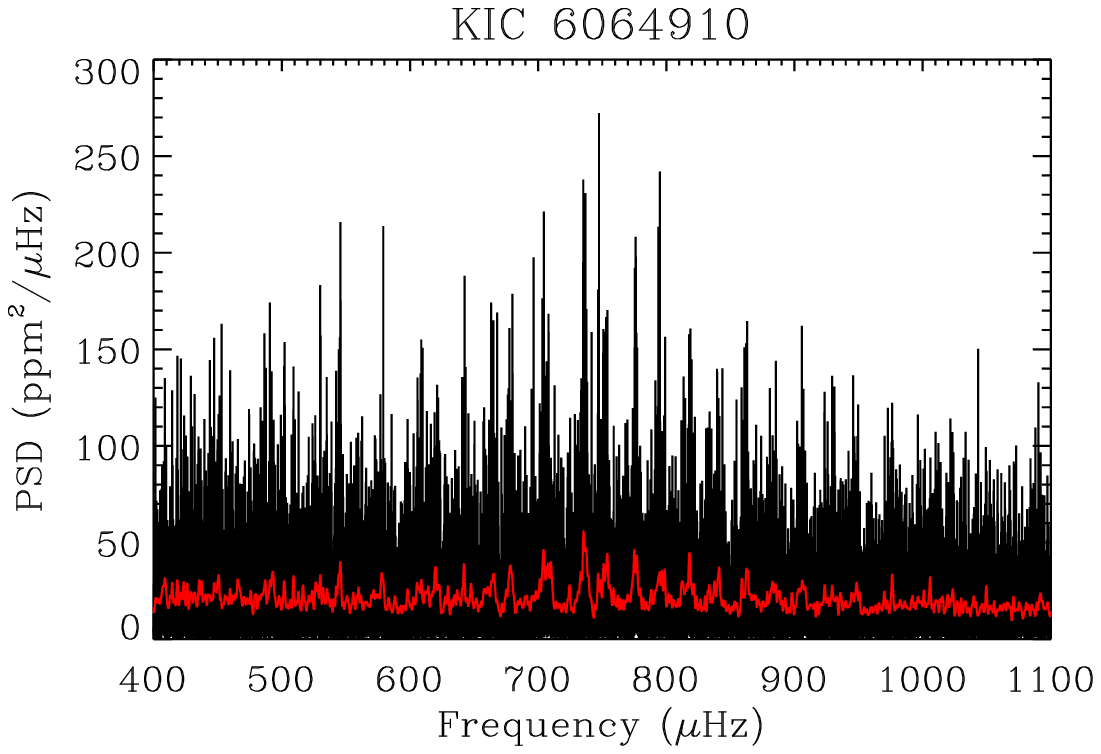}
     \includegraphics[width=7.0cm, angle=0]{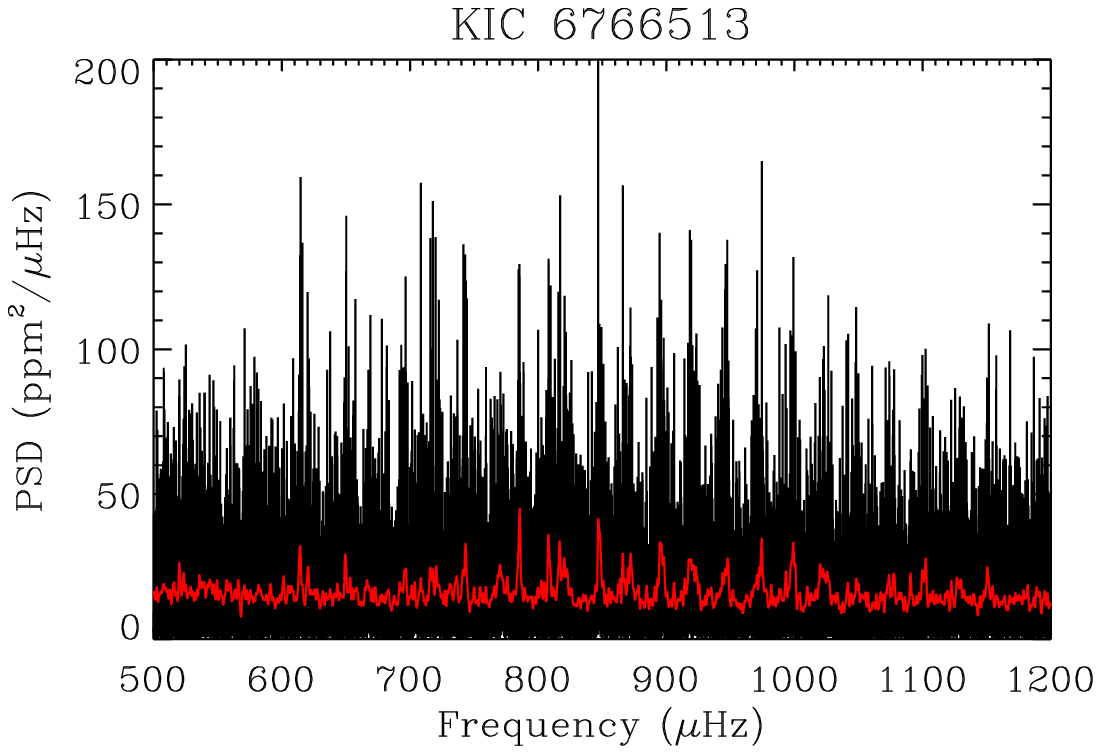}
     \includegraphics[width=7.0cm, angle=0]{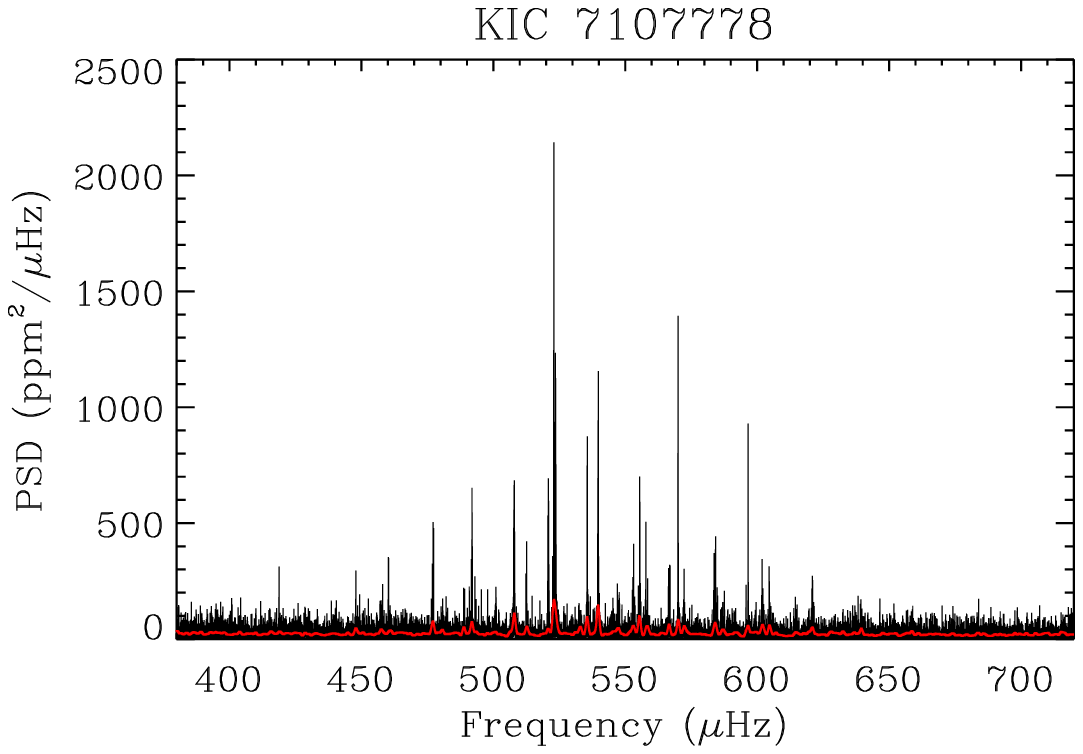}
     \includegraphics[width=7.0cm, angle=0]{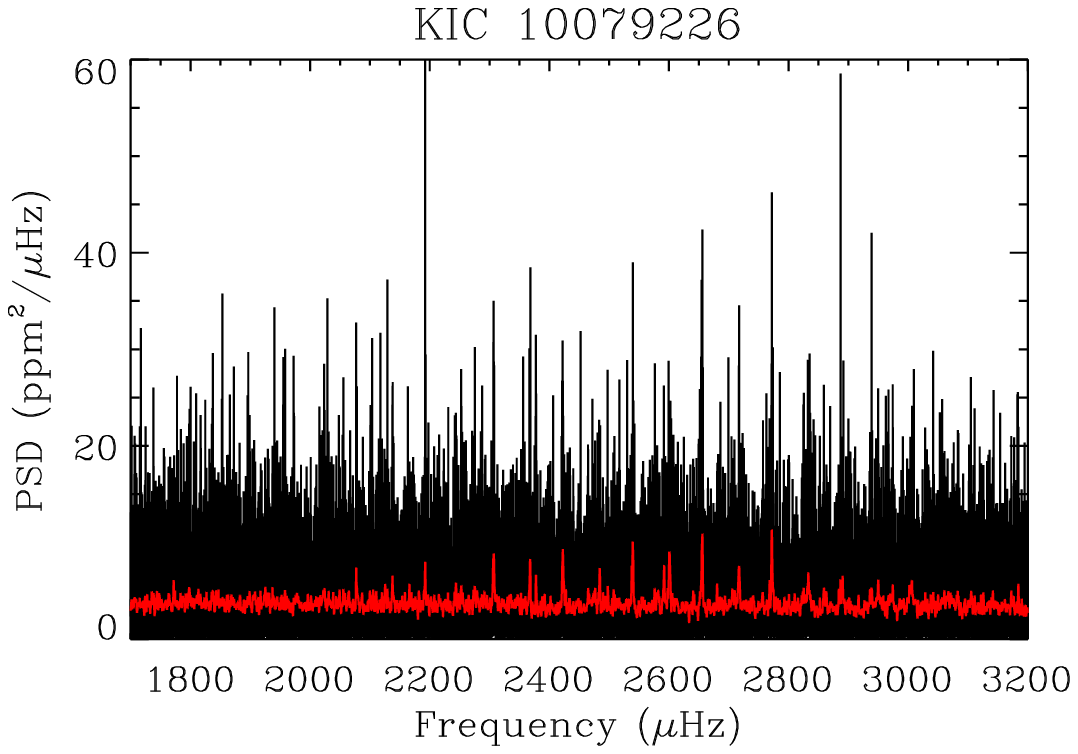}
     \includegraphics[width=7.0cm, angle=0]{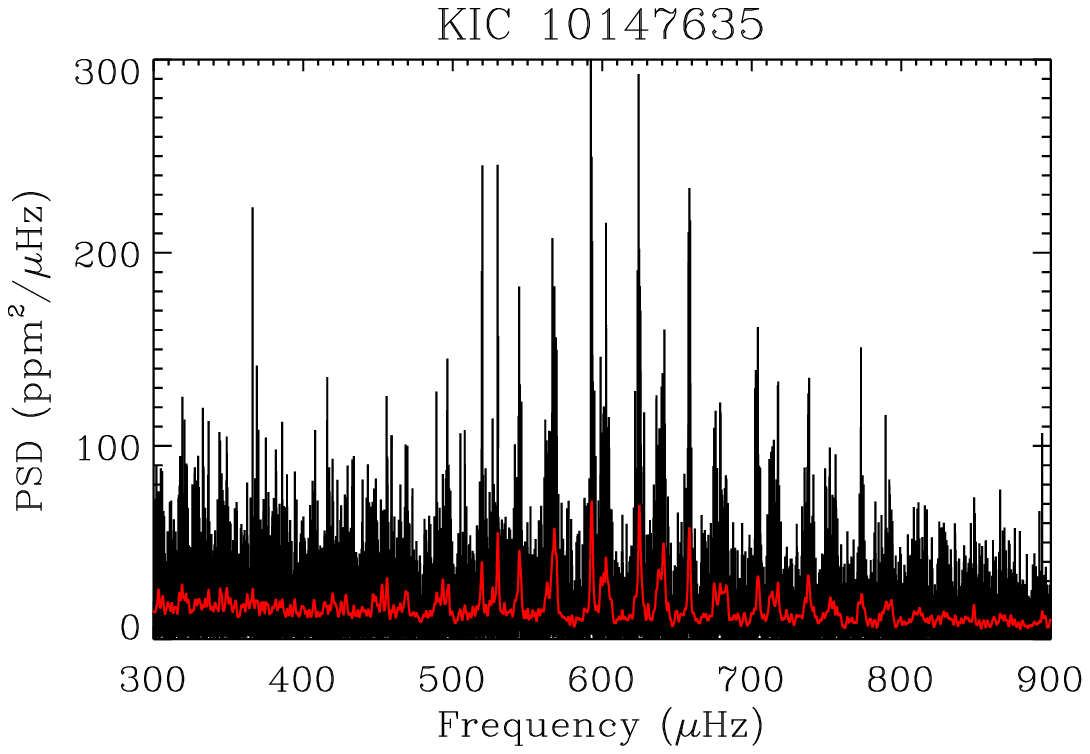}
     \includegraphics[width=7.0cm, angle=0]{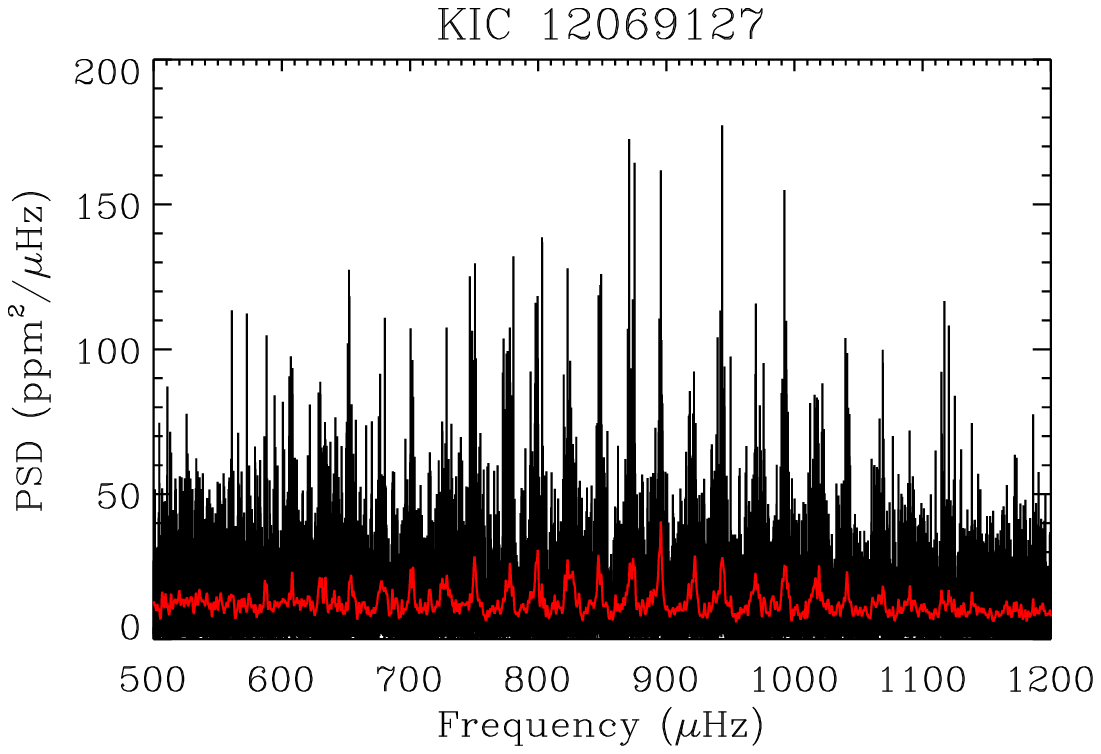}
  \caption{Power spectra of six stars. Lines in black are raw power spectrum; lines in red are smoothed ones using a Gaussian-weighted window function with width of 2$\mu$Hz.}
  \label{Fig3}
\end{figure}

Figure \ref{Fig3} displays the raw power spectra of the six targets in black and smoothed one using a Gaussian-weighted window function in red in the oscillation range. Note that the smoothed power spectra are only used to enhance the appearance, and we did not use them in the following data analysis.

\subsection{Global asteroseismic parameters}
\label{sect:aa-gap}

Mean large separation $\Delta\nu$ indicates the mean value of separations between two neighboring $l=0$ modes. It measures the pace of the increase of non-radial modes. Therefore, it can be derived utilizing the autocorrelation function (ACF) (e.g. \citealt{2006MNRAS.369.1491R}; \citealt{2014MNRAS.445.2999T}), since ACF of a series yields information about the period. ACF of $X_i,$ $i=1,...,n$ is defined as
\begin{equation}
R(k)=\frac{{\rm E}\left[(X_i-\mu)(X_{i+k}-\mu)\right]}{\sigma^2},
\end{equation}
where ``E'' is the expected value operator, and $\mu$ and $\sigma^2$ are the mean and variance respectively. The larger the ACF is, the stronger relation it shows at this specific phase $k$, which is more likely to be $\Delta\nu$. We applied the ACF method to the power spectra and found several peaks.

\begin{figure}
   \centering
  \includegraphics[width=8.0cm, angle=0]{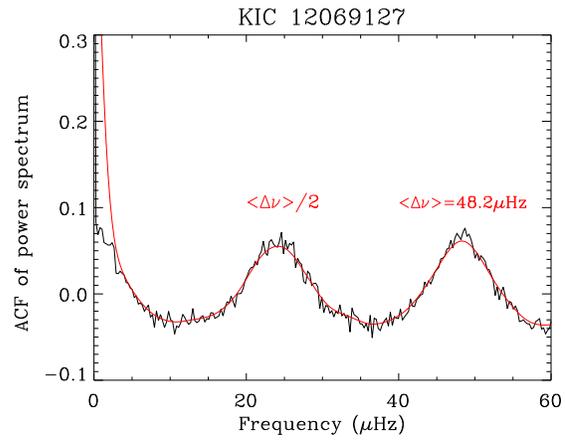}
   \caption{ACF for the power spectrum of KIC 12069127 ranging from 600 to 1100 $\mu$Hz. Black and red curves represent the original and smoothed ACF respectively. The second peak corresponds to $\Delta\nu=48.2\mu$Hz.}
   \label{Fig4}
\end{figure}

\begin{figure}
     \includegraphics[width=7.0cm, angle=0]{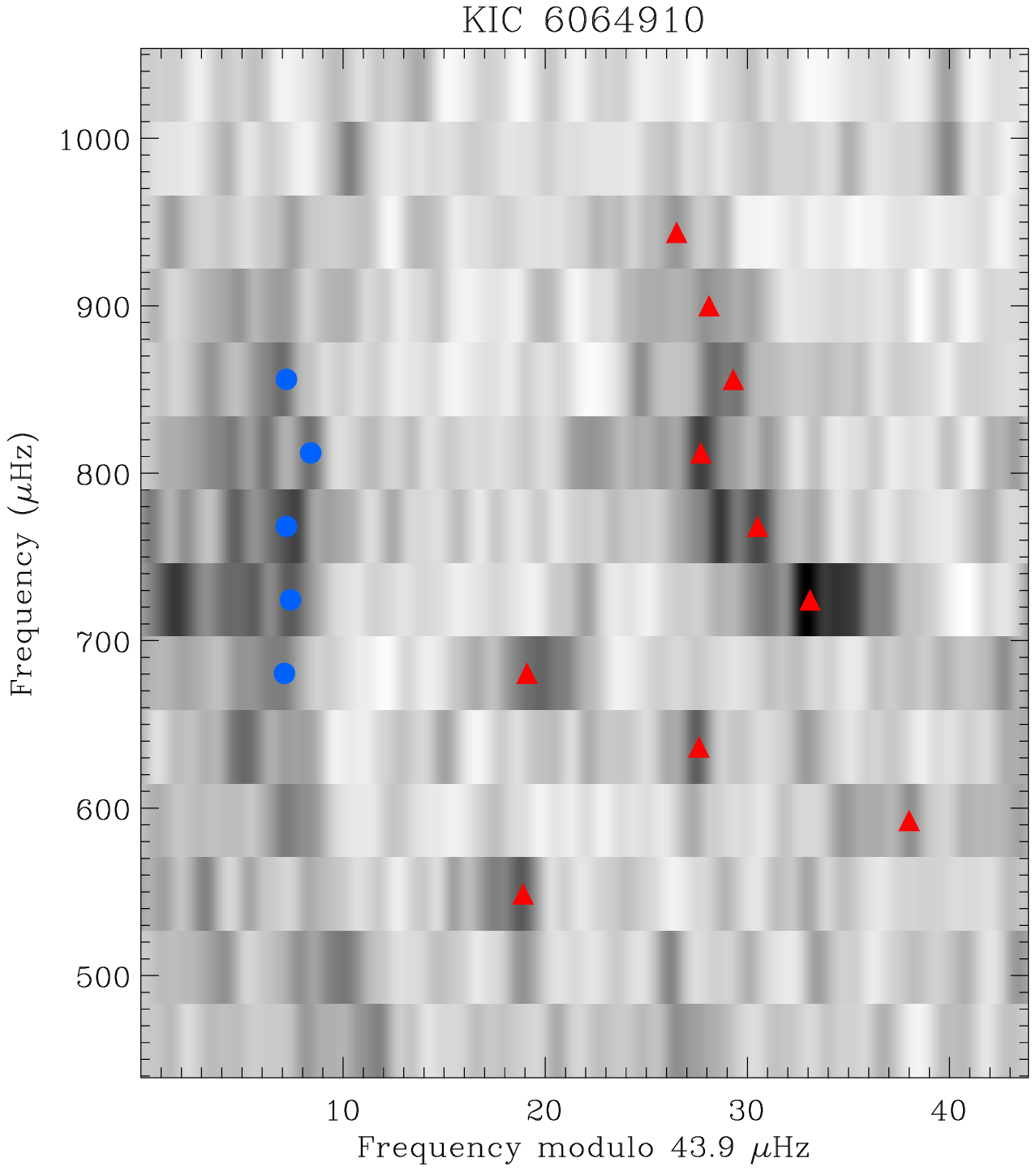}
     \hspace{1cm}
     \includegraphics[width=7.0cm, angle=0]{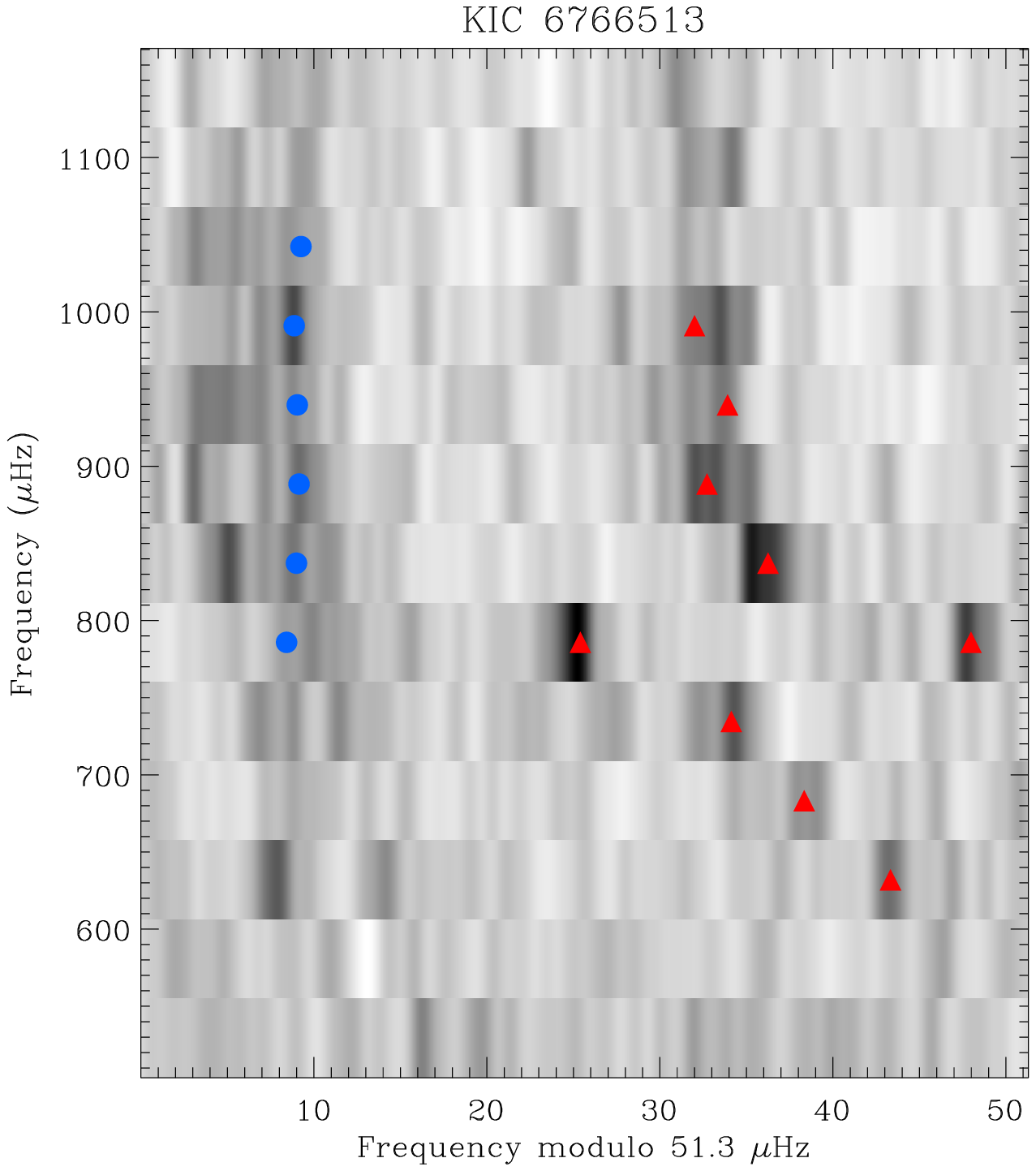}
     \includegraphics[width=7.0cm, angle=0]{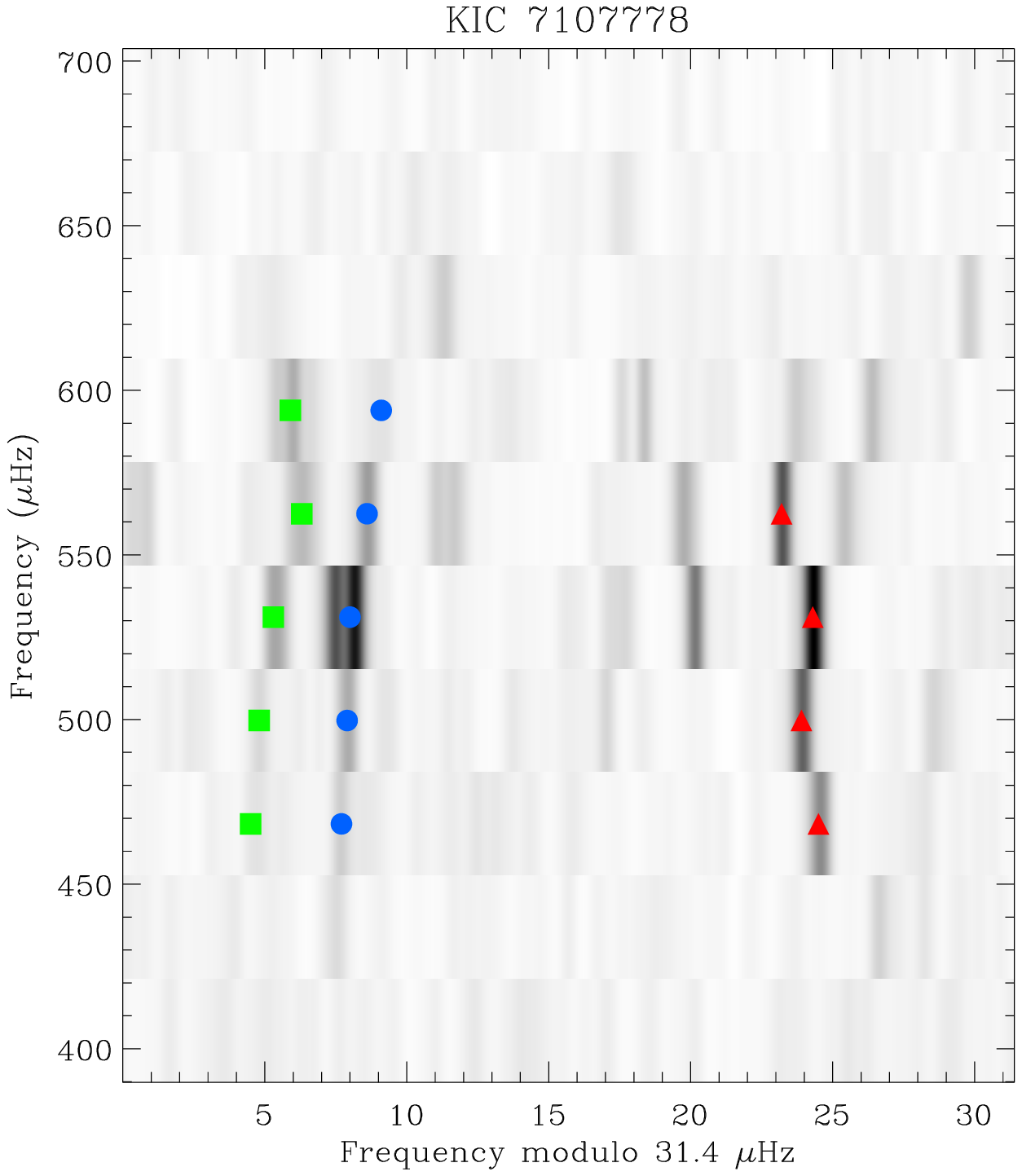}
     \hspace{1cm}
     \includegraphics[width=7.0cm, angle=0]{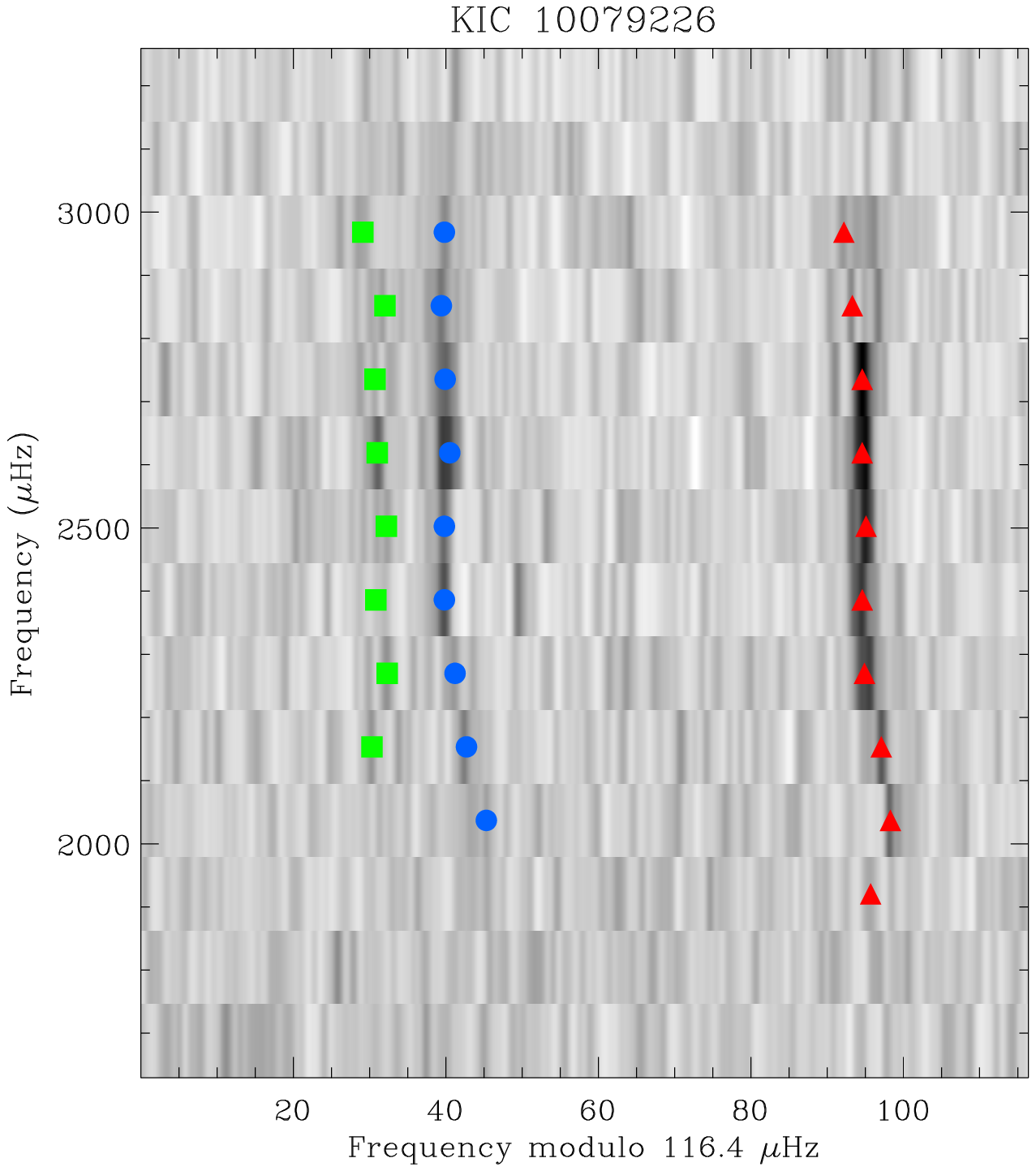}
     \includegraphics[width=7.0cm, angle=0]{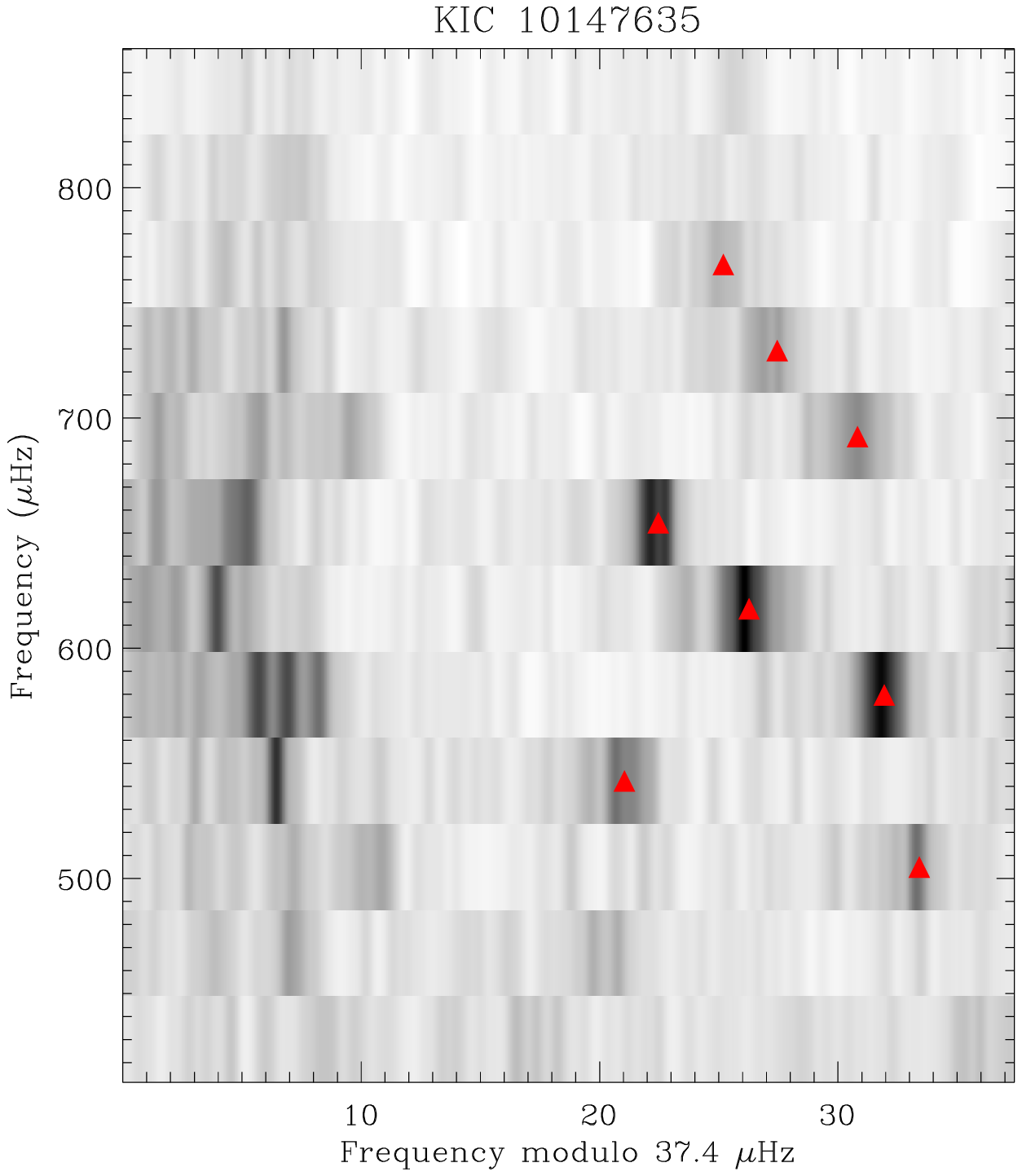}
     \hspace{1cm}
     \includegraphics[width=7.0cm, angle=0]{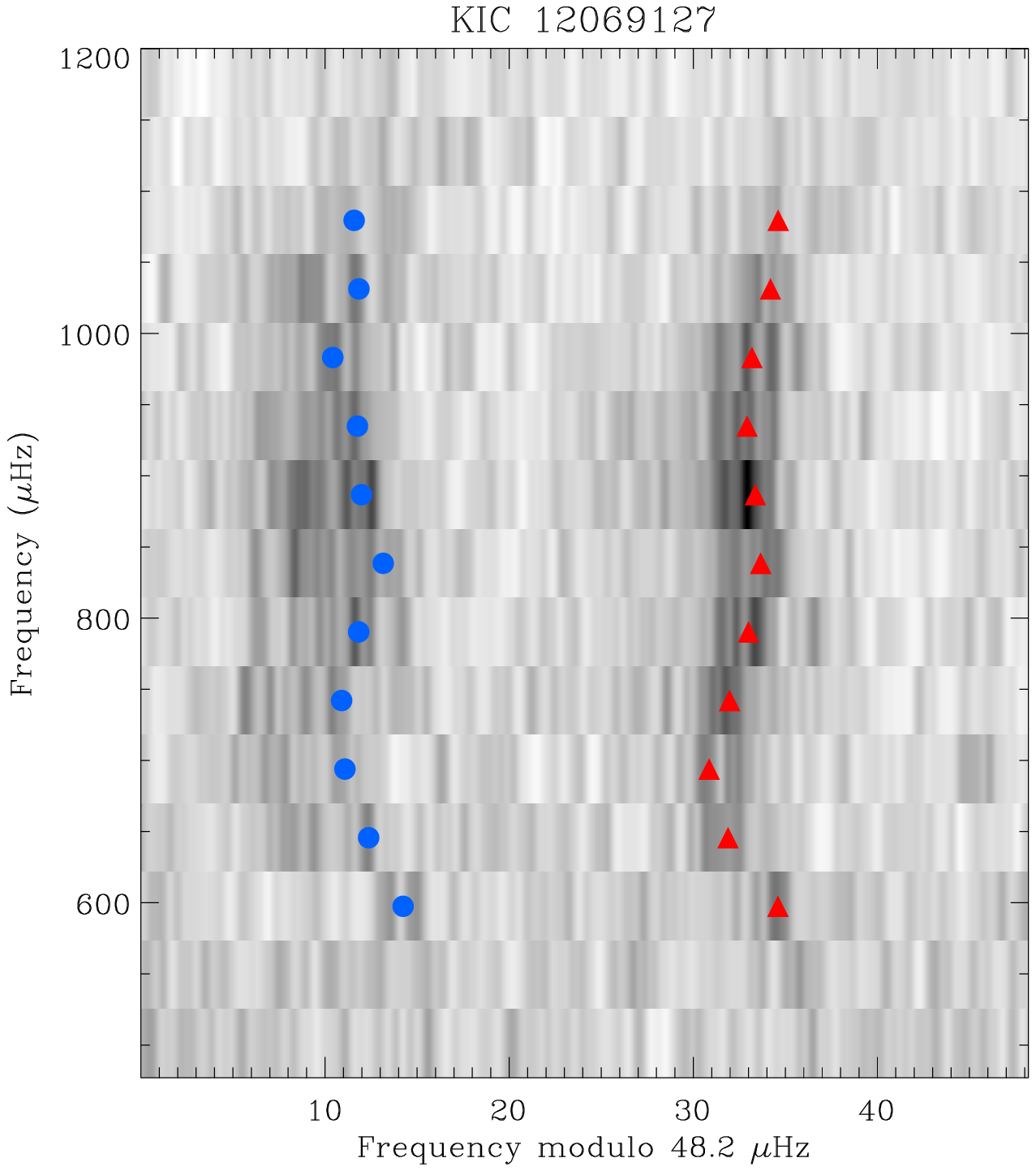}
  \caption{\'{E}chelle diagram with identified oscillation modes. Frequencies represented by circles, triangles, and squares correspond to modes with $l=$0, 1, and 2 respectively. Power spectra are shown in gray scale.}
  \label{Fig5}
\end{figure}

Figure \ref{Fig4} shows the ACF of KIC 12069127 as an example. Note that the first peak appears at $\langle\Delta\nu\rangle/2$, and it is caused by overlap between $l=0$ and $l=1$ modes. This requires us to be careful when examining our results. We validated $\Delta\nu$ with two methods. First, we took the Fourier transformation of the power spectrum in the oscillation range. A peak is expected to be found around $\Delta\nu$. Second, we used $\Delta\nu$ to plot the \'{e}chelle diagram. It extracts sections of the power spectrum in the space of $\Delta\nu$ and stacks them from bottom to top. Amplitudes of power density are displayed on a color scale. The right $\Delta\nu$ should make \'{e}chelle diagram display clear pulsation patterns. For instance, we expect to see three ridges on it for main-sequence stars, corresponding to $l=0,1,2$ modes. The patterns become more complicated when the stars evolve (discussed later), but are still recognizable. Figure \ref{Fig5} shows the \'{e}chelle diagram of six stars.

Frequency of maximum power $\nu_{\rm{max}}$ measures the location of power excess. It can be obtained by heavily smoothing the power spectrum and the central frequency is denoted by $\nu_{\rm{max}}$ (\citealt{2008A&A...485..813C}; \citealt{2009MNRAS.400L..80S}). Here, we checked this result by collapsed ACF, according to \cite{2009CoAst.160...74H}. First, we split the power spectrum into the same logarithmic bins, and smooth them with a median filter as the background of the power spectrum. Second, we subtract the background derived above from the raw power spectrum, and divide the residual power spectrum into subsets roughly equal to 4$\Delta\nu$ with an overlap. We calculated ACF for each subset. Third, we collapse the ACF of each subset over all frequency spacing. We fit the collapsed ACF with a Gaussian profile and its frequency of maximum value is believed to be $\nu_{\rm{max}}$.

\begin{figure}
   \centering
  \includegraphics[width=8.0cm, angle=0]{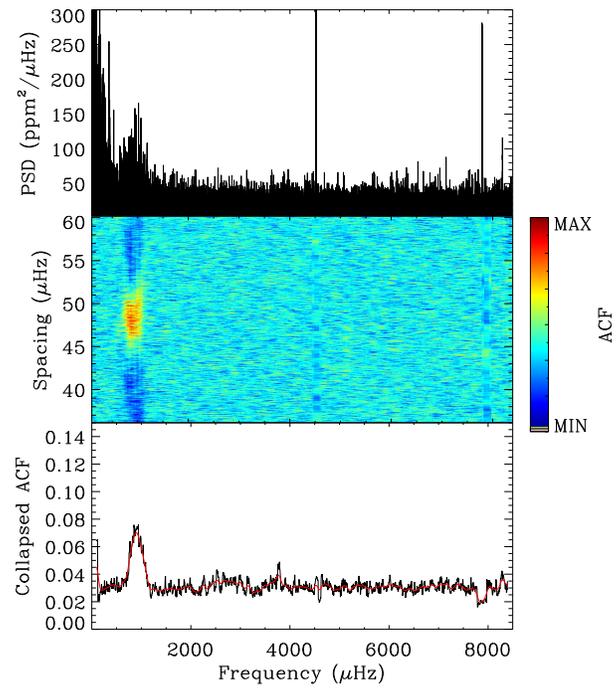}
   \caption{\emph{Top panel}: the original power spectrum of KIC 12069127. \emph{Middle panel}: ACF displayed on a color scale of each subset for each spacing. \emph{Bottom panel}: collapsed ACF in black and smoothed one in red.}
   \label{Fig7}
\end{figure}

Figure \ref{Fig7} shows the power spectrum of KIC 12069127 in the top panel, together with ACF of each subset for each spacing in the middle panel, and collapsed ACF in the bottom panel.

After we checked our results with different approaches for quality assurance, we compared our results, both $\Delta\nu$ and $\nu_{\rm{max}}$, to previous literature. \cite{2014ApJS..210....1C} only used the first ten months of \emph{Kepler} data, and for the case of our six targets, only one month SC data was used.

Table \ref{tab:dnu+numax} shows results from the two works. We note that $\Delta\nu$ shows good accordance, while $\nu_{\rm{max}}$ has slight deviations. This may be due to two reasons. First, we used different data, as illustrated above. Second, the definition of $\nu_{\rm max}$ is slightly ambiguous, so that it carries an intrinsic uncertainty. Furthermore, the choices of different smoothing function may enhance the uncertainty. Because the differences are not that large and $\nu_{\rm{max}}$ does not show strong constraints on the following stellar models, we then omitted them.

\subsection{Oscillation Frequencies}
\label{sect:aa-of}

In order to excavate deeper seismic information, we extracted oscillaion frequencies of the six targets. We started by reviewing characteristics of solar-like oscillation. The signature of p modes can be well described using asymptotic theory controlled by radial order $n$ and angular degree $l$. The approximate expression can be written as \citep{1980ApJS...43..469T}:
\begin{equation}
\nu_{nl}=\Delta\nu(n+\frac{l}{2}+\epsilon)+l(l+1)D_0,
\end{equation}
where coefficient $\epsilon$ and $D_0$ depend on stellar conditions. This expression indicates p mode frequencies are equally spaced in frequency (i.e. $\Delta\nu$). The large separation is related to the sound radius through:
\begin{equation}
\Delta\nu=(2\int_{0}^{R}\frac{dr}{c})^{-1},
\end{equation}
and $c$ and $R$ are the sound speed and the stellar radius, respectively. $\Delta\nu$ is related to the acoustic diameter. Higher order of g modes can also be described by an asymptotic relation \citep{1980ApJS...43..469T}:
\begin{equation}
\Pi_{nl}=\nu_{nl}^{-1}\simeq\Delta\Pi_l(n+\epsilon_g),
\end{equation}
where
\begin{equation}
\Delta\Pi_l=\frac{2\pi^2}{\surd{l(l+1)}}(\int_{r_1}^{r_2}N\frac{dr}{r})^{-1},
\end{equation}
where $N$ is the buoyancy frequency which controls the behavior of g modes and is given by
\begin{equation}
N^2=g(\frac{1}{\Gamma_1}\frac{\rm{d}\ln{p}}{\rm{d}{r}}-\frac{\rm{d}\ln{\rho}}{\rm{d}{r}}).
\end{equation}
This indicates g mode frequencies are placed equally spaced in period (i.e. $\Delta\Pi$).

Several methods to extract oscillation frequencies in a global way have been put forward, for example, Bayesian Markov Chain Monte Carlo (\citealt{2011A&A...527A..56H}; \citealt{2008CoAst.157...98B}), Maximum Likelihood Estimation (\citealt{1998A&AS..132..107A}); however, when the power spectrum reveal p and g mixed modes with low S/N, global analysis is not advantageous, because there exist several frequencies that are hard to determine, and it is easy for an automatic program to wrongly determine these frequencies. Therefore, here we derived them separately based on asymptotic theory and visual inspection. The identified modes are fitted with Lorentzian profiles using the least-squares minimization. The Lorentzian model is:
\begin{equation}
L_i=\frac{A}{\left(\frac{\nu_i-\nu_0}{\Gamma}\right)^2+1},
\end{equation}
where the three free parameters are amplitude $A$, frequency centroid $\nu_0$, and linewidth $\Gamma$. We regard $\nu_0$ as the oscillation frequency. The objective function is written as,
\begin{equation}
Q=\sum^{n}_{i=1}\left(P_i-L_i\right)^2,
\end{equation}
where $P_i$ is the power spectrum density. The objective function can be minimized by taking partial derivatives and setting to zero, i.e.
\begin{equation}
\begin{aligned}
\frac{\partial{Q}}{\partial{A}}=0,\\
\frac{\partial{Q}}{\partial{\nu_0}}=0,\\
\frac{\partial{Q}}{\partial{\Gamma}}=0.
\end{aligned}
\end{equation}
Values for frequency centroid $\nu_0$ are derived by solving the above set of non-linear equations.

We present individual mode frequencies $\nu_0$ in Table \ref{frequency}. For KIC 7107778 and KIC 10079226, we obtained their modes $\nu_{nl}$ with $l=0,1,2$. For KIC 6064910, KIC 6766513 and KIC 12069127, we obtained their modes $\nu_{nl}$ with $l=0,1$. $l=2$ modes of those stars are difficult to identify due to low S/N. For KIC 10147635, only modes $\nu_{nl}$ with $l=1$ were extracted, because the $l=0,2$ modes were significantly mixed. In figure \ref{Fig5}, We mark the identified modes with circles, triangles, and squares for $l=0, 1,$ and $2$, respectively.

\section{Grid modeling}
\label{sect:grid}
Some stellar fundamental parameters, e.g. $M$, $R$, can be directly deduced by seismic parameters (discussed below). However, to further investigate and analyze the six targets comprehensively, we constructed stellar grid models. The main theme of grid modeling is to construct models in a large range and select models which meet the constraints, including seismic constraints (e.g. $\Delta\nu$ and $\nu_{\rm{max}}$) and non-seismic constraints (e.g. stellar atmospheric parameters). The properties of these qualified models are regarded as properties of the stars.

\subsection{Modeling parameters and input physics}

We computed stellar models with the Yale Rotating Stellar Evolution Code (\citealt{2008Ap&SS.316...31D}; \citealt{1990BAAS...22..746P}; \citealt{1992ApJS...78..179P}). The input parameter, mass $M$ was estimated with scaling relations. Mean large separation $\Delta\nu$ is related to mean density of the star, i.e. $\Delta\nu\propto\sqrt{\rho}$, and the frequency of maximum power $\nu_{\rm{max}}$ is related to both surface gravity and effective temperature of the star, i.e. $\nu_{\rm{max}}\propto{g/\sqrt{T_{\rm{eff}}}}$ (e.g. \citealt{1993ASPC...42..347C} and \citealt{1995A&A...293...87K}). Hence, $\Delta\nu$ and $\nu_{\rm{max}}$ can be expressed in terms of the solar values:
\begin{equation}
\frac{\Delta\nu}{\Delta\nu_\odot}\approx(\frac{M}{M_\odot})^{1/2}(\frac{R}{R_\odot})^{-3/2},
\end{equation}
\begin{equation}
\frac{\nu_{\rm{max}}}{\nu_{\rm{max},\odot}}\approx(\frac{M}{M_\odot})(\frac{R}{R_\odot})^{-2}(\frac{T_{\rm{eff}}}{T_{\rm{eff},\odot}})^{-1/2},
\end{equation}
where $\Delta\nu_\odot=135.1$ $\mu\rm{Hz}$, $\nu_{\rm{max},\odot}=3050$ $\mu\rm{Hz}$ \citep{2013ARA&A..51..353C} and $T_{\rm{eff,\odot}}=5777$ K. Combining the two equations above, we obtained
\begin{equation}
\frac{M}{M_\odot}\approx(\frac{\Delta\nu}{\Delta\nu_\odot})^{-4}(\frac{\nu_{\rm{max}}}{\nu_{\rm{max},\odot}})^{3}(\frac{T_{\rm{eff}}}{T_{\rm{eff},\odot}})^{3/2}.
\end{equation}
This requires the grid should at least cover $0.8\sim1.8$ $M_\odot$, considering uncertainties associated with each parameter. Spectroscopic observation requires $\rm{[Fe/H]}$ of the grid range from $-0.3\sim0.4$ $\rm{dex}$. We ignored convection overshooting and treated convection with standard mixing-length theory \citep{1958ZA.....46..108B}, with three mixing parameters 1.75, 1.84 and 1.95. In particular, 1.84 originates from the solar calibrated model of YREC (see \citealt{2016arXiv160905707W}). The free input parameters mass $M$, initial metallicity [Fe/H] and mixing length parameter $\alpha$, are summarized in Table \ref{tab3}.

The input physics is set as follows. We adopted NACRE Nuclear reaction rates in \cite{1995RvMP...67..781B}, equation of state tables in \cite{2002ApJ...576.1064R}, OPAL high-temperature opacities in \cite{1996ApJ...464..943I}, and low-temperature opacities in \cite{2005ApJ...623..585F}. Atomic diffusion was considered only under initial masses $<1.1$ $M_{\odot}$, with the formulation of \cite{1994ApJ...421..828T}. The element abundance ratio was estimated by
\begin{equation}
\rm{[Fe/H]}=\log{\rm{(Z/X)}}-\log{(\rm{Z/X})_{\odot}},
\end{equation}
where $\rm{(Z/X)}_{\odot}=0.0231$ \citep{1998SSRv...85..161G}. We treated the initial helium abundance of these models as
\begin{equation}
\rm{Y}_{ini}=0.248+\rm{Z}_{ini}\cdot\frac{\Delta \rm{Y}}{\Delta \rm{Z}},
\end{equation}
where ${\Delta \rm{Y}}/{\Delta \rm{Z}}=1.33$, which also comes from a solar calibrated model \citep{2016arXiv160905707W}. All models were calculated from Hayashi lines to red giant branch.

\subsection{Constraining models}

We selected qualified models of the six targets, which match the requirements imposed by observational constraints: $T_{\rm{eff}}$, $\rm{[Fe/H]}$, $\Delta\nu$ and $\nu_{\rm{max}}$ (Table \ref{tab1} and \ref{tab:dnu+numax}). The qualifications were estimated with a Bayesian approach (\citealt{2010ApJ...710.1596B}; \citealt{2010A&A...522A...1K}).

We assigned an overall probability for each model $M_i$, with respect to posterior probability $I$ and observations $D$. According to Bayes' theorem,
\begin{equation}
p\left(M_i|D,I\right)=\frac{p\left(M_i|I\right)p\left(D|M_i,I\right)}{p\left(D|I\right)}.
\end{equation}
The prior probability is set to a uniform value:
\begin{equation}
p\left(M_i|I\right)=\frac{1}{N_m},
\end{equation}
with $N_m$ being the number of models. The likelihood is expressed as,
\begin{equation}
p\left(D|M_i,I\right)={\mathcal L}\left(T_{\rm eff},{\rm [Fe/H]},\Delta\nu,\nu_{\rm max}\right)={\mathcal L}_{T_{\rm eff}}{\mathcal L}_{\rm [Fe/H]}{\mathcal L}_{\Delta\nu}{\mathcal L}_{\nu_{\rm max}},
\end{equation}
where
\begin{equation}
{\mathcal L}_{T_{\rm eff}}=\frac{1}{\sqrt{2\pi}\sigma_{T_{\rm eff}}}\exp\left[\frac{-\left(T_{\rm eff,obs}-T_{\rm eff,model}\right)^2}{2\sigma^2_{T_{\rm eff}}}\right],
\end{equation}
\begin{equation}
{\mathcal L}_{\rm [Fe/H]}=\frac{1}{\sqrt{2\pi}\sigma_{\rm [Fe/H]}}\exp\left[\frac{-\left({\rm [Fe/H]_{obs}}-{\rm [Fe/H]_{model}}\right)^2}{2\sigma^2_{\rm [Fe/H]}}\right],
\end{equation}
\begin{equation}
{\mathcal L}_{\Delta\nu}=\frac{1}{\sqrt{2\pi}\sigma_{\Delta\nu}}\exp\left[\frac{-\left(\Delta\nu_{\rm obs}-\Delta\nu_{\rm model}\right)^2}{2\sigma^2_{\Delta\nu}}\right],
\end{equation}
\begin{equation}
{\mathcal L}_{\nu_{\rm max}}=\frac{1}{\sqrt{2\pi}\sigma_{\nu_{\rm max}}}\exp\left[\frac{-\left(\nu_{\rm max,obs}-\nu_{\rm max,model}\right)^2}{2\sigma^2_{\nu_{\rm max}}}\right],
\end{equation}
with subscript ``obs'' and ``model'' being observation and model values respectively. The four parameters that compose the likelihood were measured independently, so the multiplication makes sense. Note that the model values of $\Delta\nu$ and $\nu_{\rm max}$ are derived with a scaling relation. The normalization factor can be added from each model probability
\begin{equation}
p\left(D|I\right)=\sum_{j=1}^{N_m}p\left(M_j|I\right)p\left(D|M_j,I\right).
\end{equation}
By canceling out the constant prior probabilities, the Bayes' theorem simplifies to
\begin{equation}
p\left(M_i|D,I\right)=\frac{p\left(D|M_i,I\right)}{\sum_{j=1}^{N_m}p\left(D|M_j,I\right)}.
\end{equation}
The above equation is used to derive posterior probability for each model. By constructing the marginal probability distribution of each parameter, we estimated their values and assigned a 1$\sigma$ deviation from median values as the uncertainties.

The oscillation patterns revealed from the corresponding '{e}chelle diagram, i.e. figure \ref{Fig5}, accompanied by the identified oscillation frequencies $\nu_{nl}$, indicate the evolutionary stage of these stars. The plot for KIC 10079226 have three nearly perpendicular ridges, suggesting it is a main-sequence star with p mode oscillations. KIC 12069127 is also ``main-sequence like''. Evolved stars turn out to show p-g mixed modes for $l>0$. An identifying signature is that $l=1$ modes deviate from the ridge, and overlap to form a slope, for example, KIC 6064910, KIC 6766513 and KIC 10147635. Additionally, KIC 7107778 displays a more sophisticate pattern, partly because it is more evolved than the others.

\begin{figure}
\centering
     \includegraphics[width=7.0cm, angle=0]{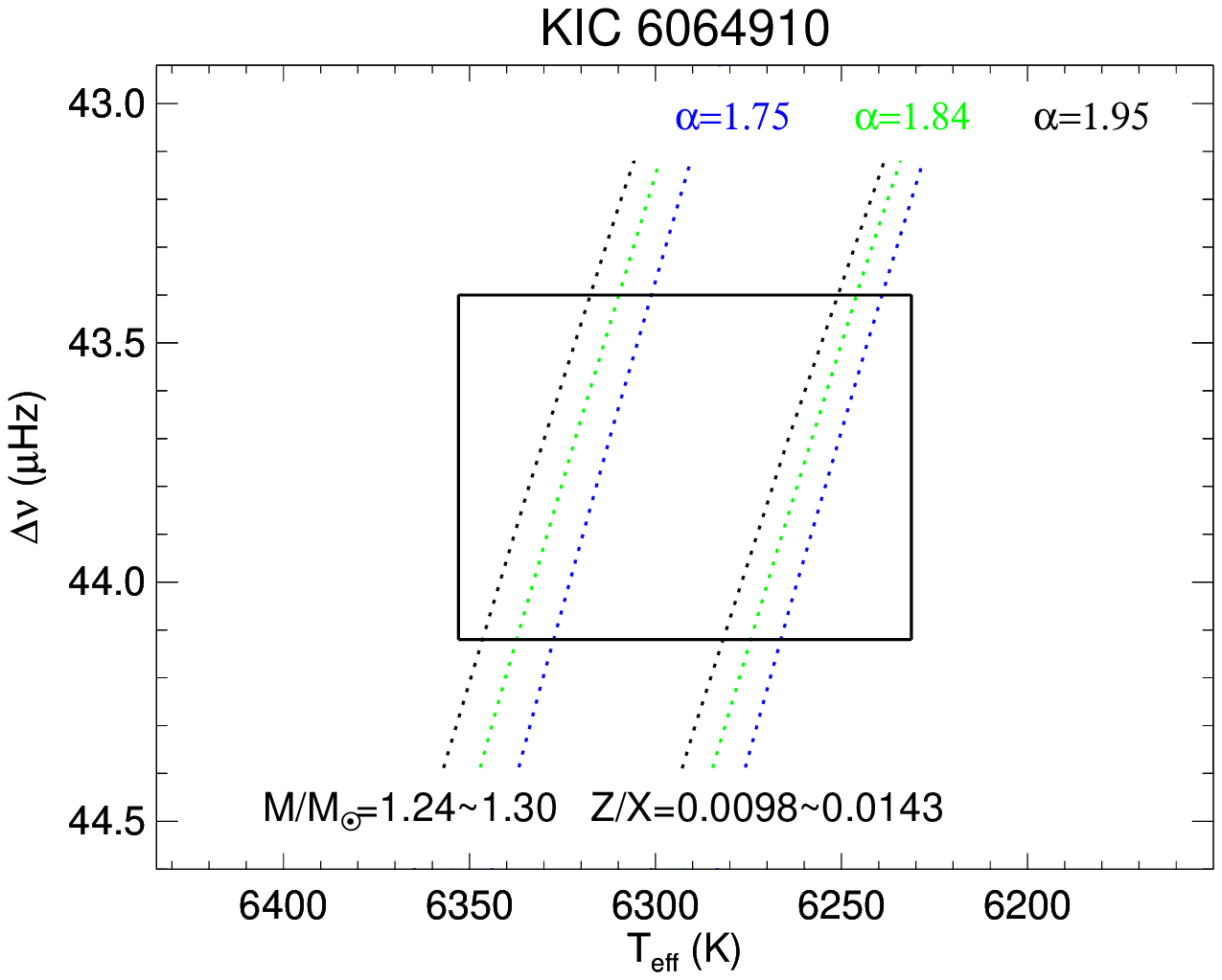}
     \includegraphics[width=7.0cm, angle=0]{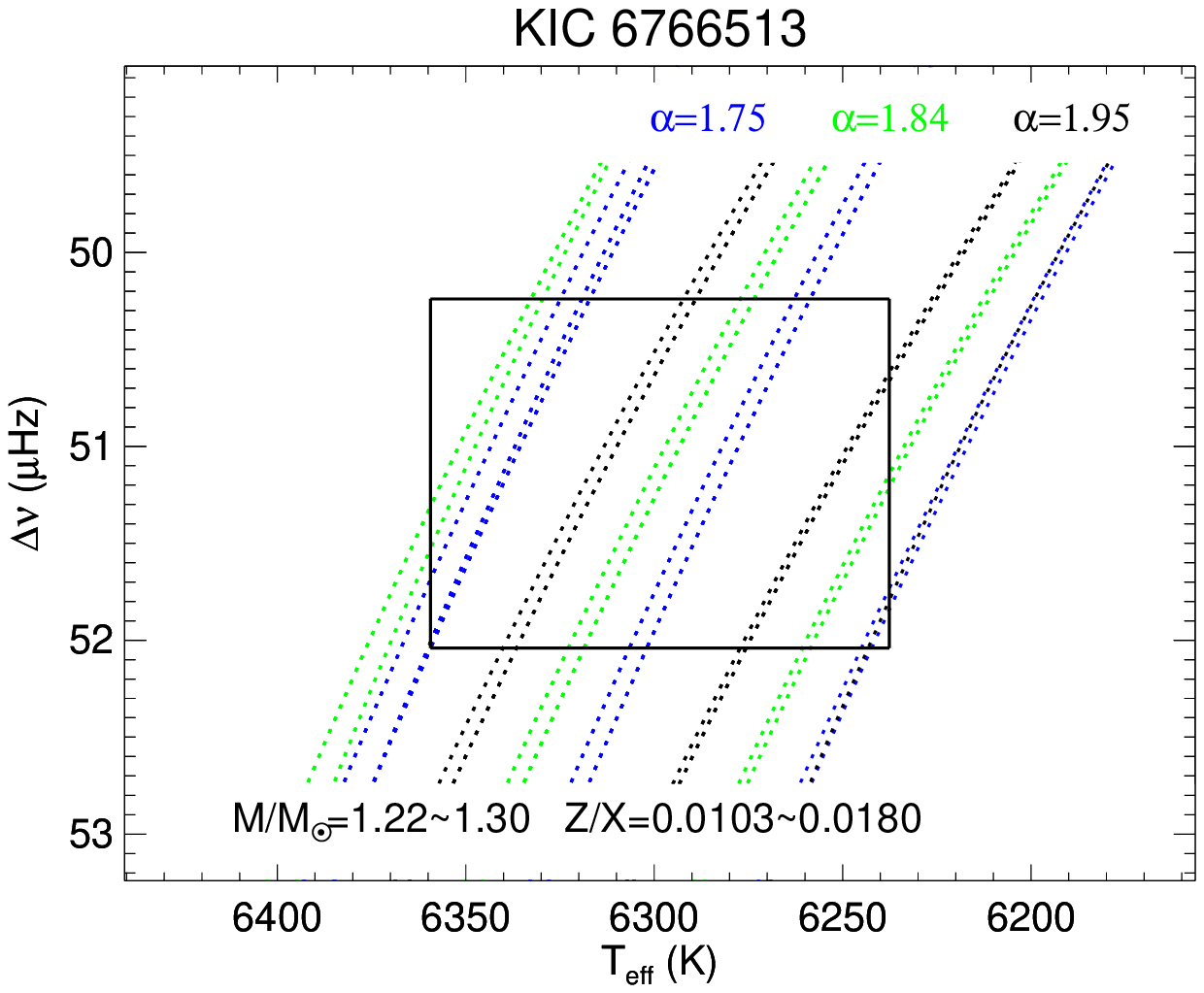}
     \includegraphics[width=7.0cm, angle=0]{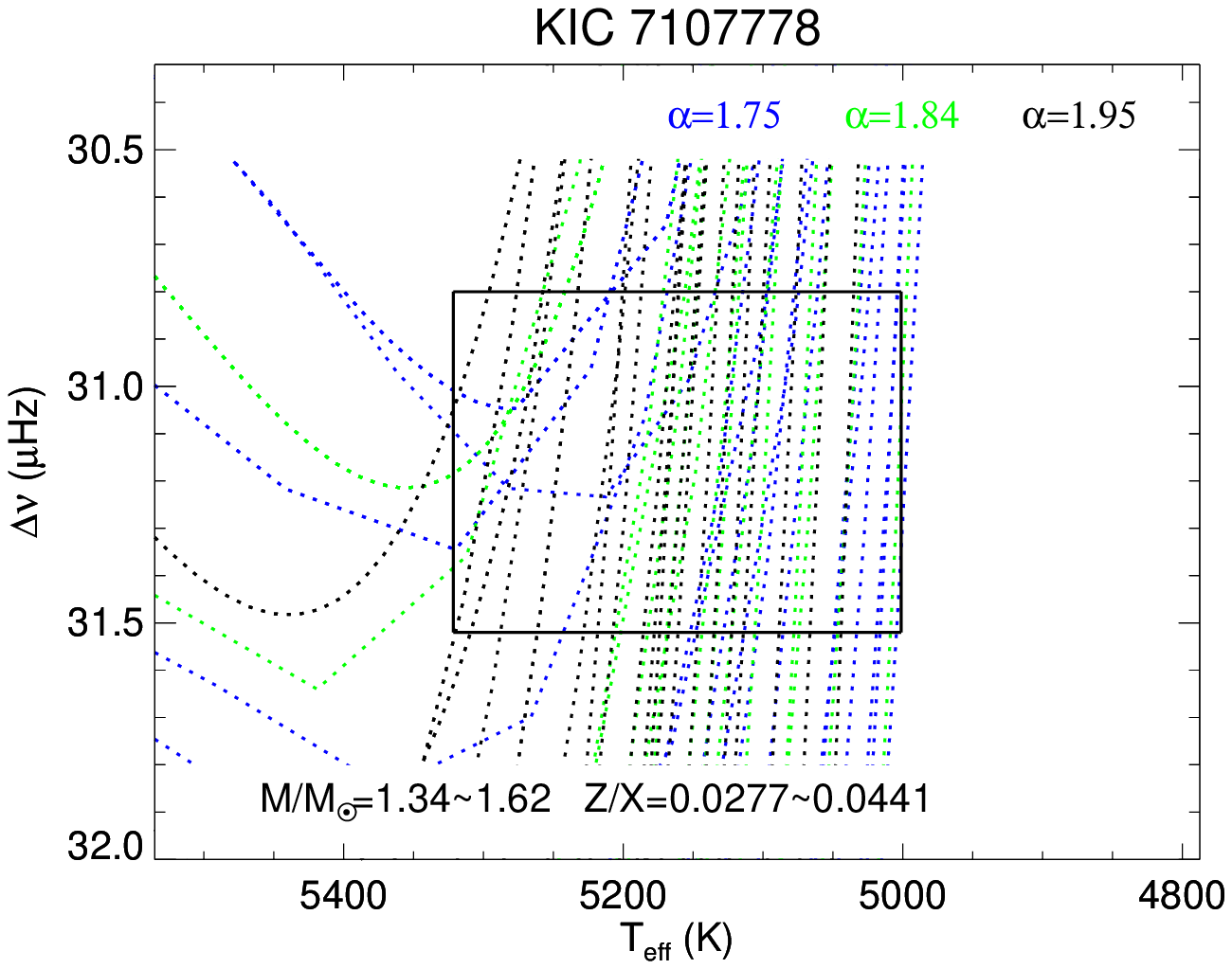}
     \includegraphics[width=7.0cm, angle=0]{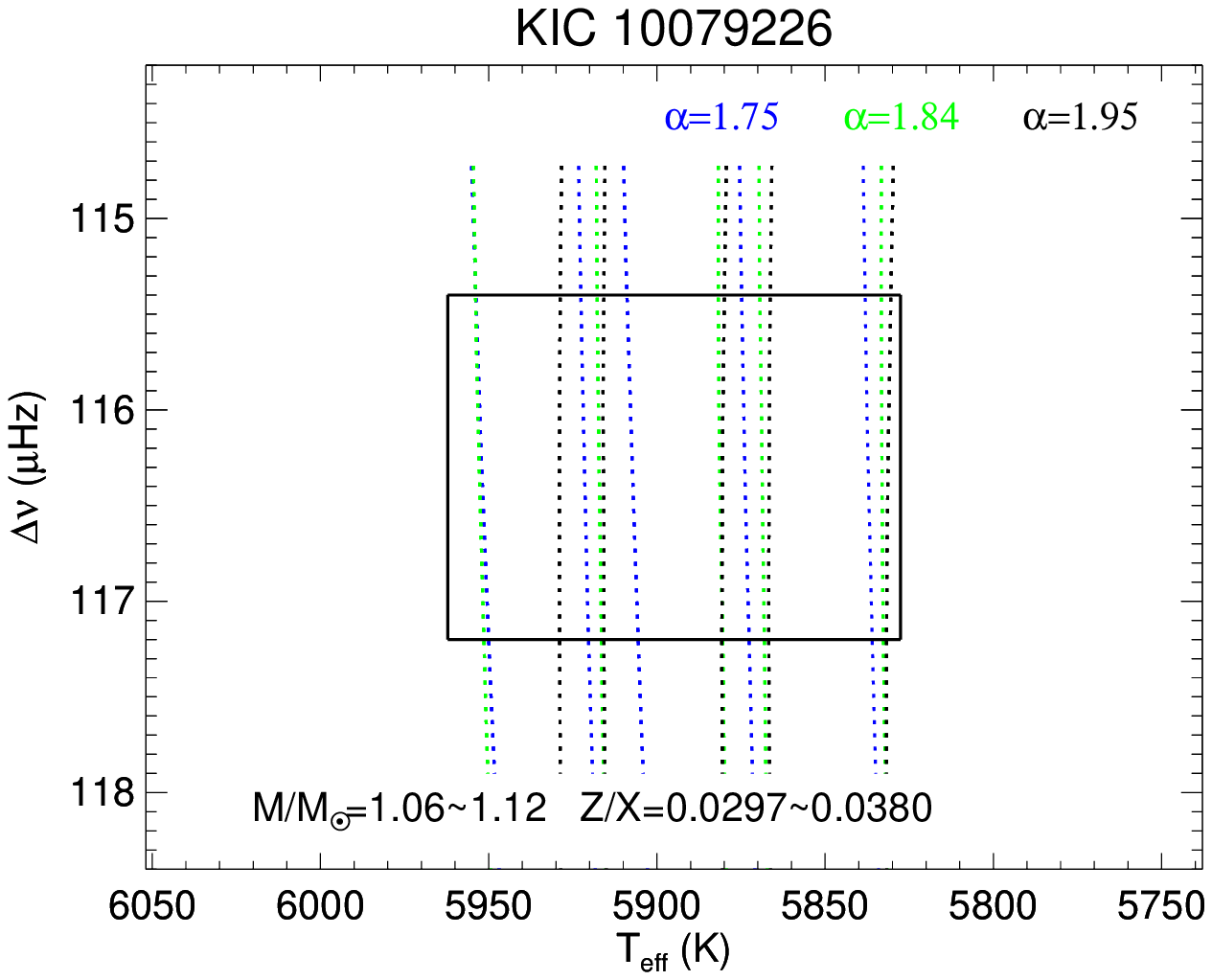}
     \includegraphics[width=7.0cm, angle=0]{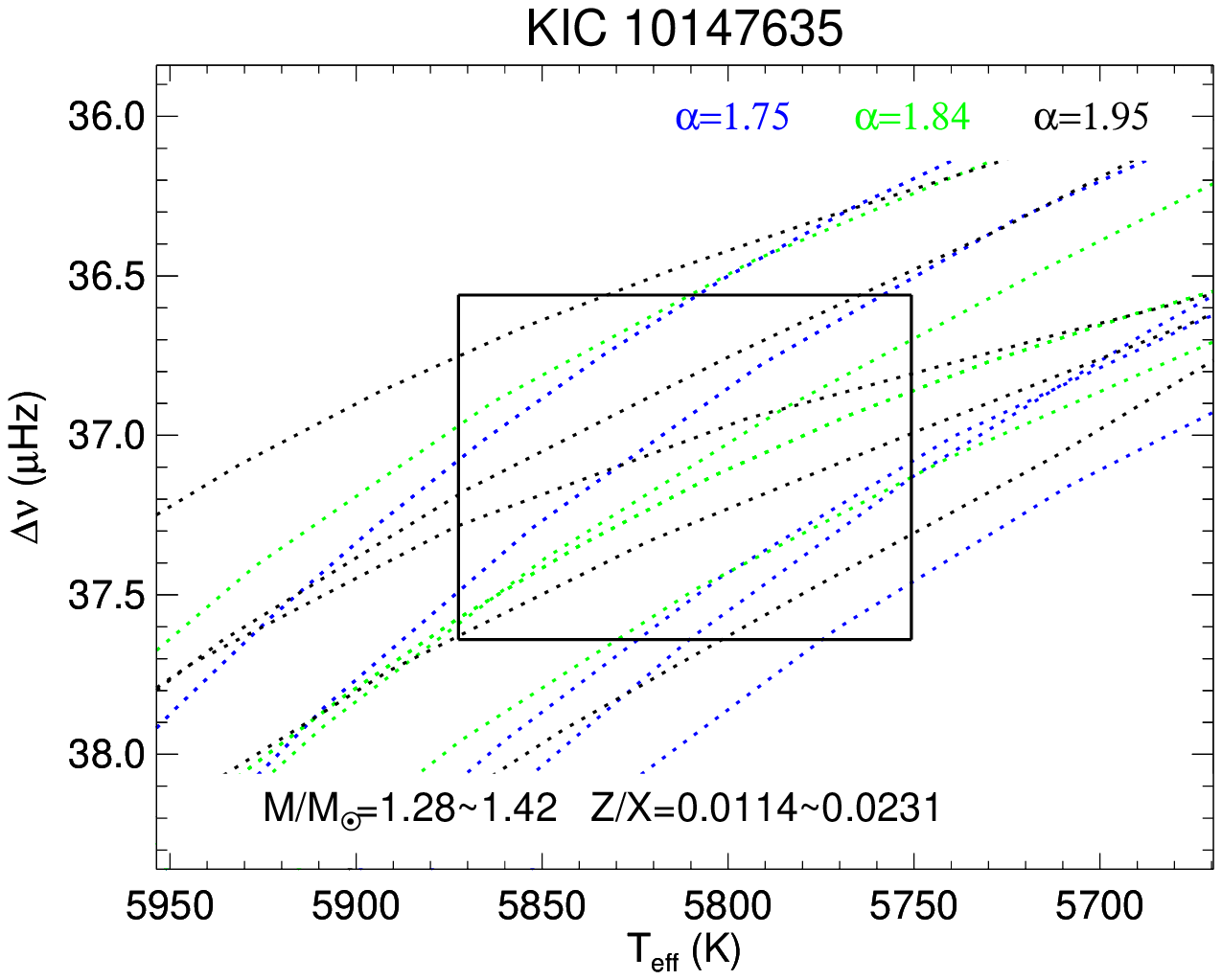}
     \includegraphics[width=7.0cm, angle=0]{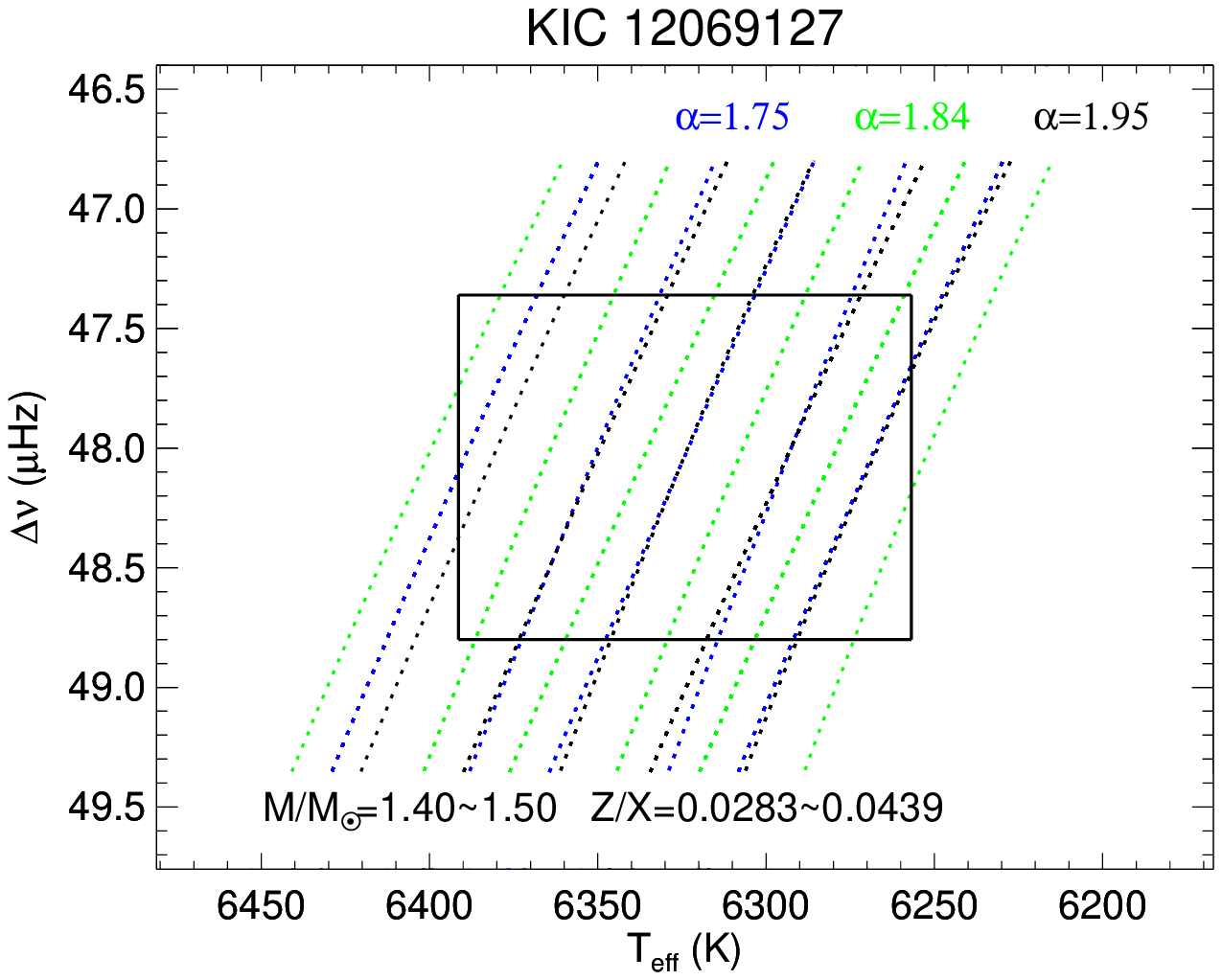}
  \caption{Models for each star on the $\Delta\nu$-$T_{\rm eff}$ diagram. The black solid squares are error boxes, representing constraints on $T_{\rm eff}$, [Fe/H], $\Delta\nu$ and $\nu_{\rm max}$. Models delineated by blue, green and black lines correspond to mixing length parameter 1.75, 1.84, and 1.95, respectively. Estimated mass $M$ and abundance ratio $Z/X$ for each star are shown at the bottom of each diagram.}
  \label{Fig1}
\end{figure}

Figure \ref{Fig1} shows the evolution tracks with models falling into error boxes (black solid squares) on the $\Delta\nu$-$T_{\rm eff}$ diagram. The error boxes are multidimensional (but are displayed as a plane here), representing constraints from $\Delta\nu$, $\nu_{\rm max}$, $T_{\rm eff}$ and [Fe/H]. Models with mixing length parameter 1.75, 1.84, and 1.95 are displayed in blue, green and black dotted lines. At the main-sequence stage, the star will go through a rather stable process when the temperature changes little with $\Delta\nu$ decreasing. The sub-giant star experiences a rapid temperature decline, and with $\Delta\nu$ decreasing. A red giant star is stably burning the hydrogen shell with an equable temperature, and $\Delta\nu$ continues to decrease. Therefore, we note that KIC 10079226 is still in the main-sequence stage, and the rest of them are sub-giant stars. According to our previous analysis on oscillation behavior, KIC 12069127 is ``main-sequence like'', but here it turns out to be a sub-giant star. This is because it is still at the early stage of a sub-giant, when p modes are not quite affected by g modes, so that the \emph{``avoided crossing''} effects are weaker. The other stars are in good accordance with their oscillation behavior on the \'{e}chelle diagram.

Table \ref{tab4} presents model parameters estimated by the Bayesian approach, together with those estimated by \cite{2014ApJS..210....1C}. Results from the two studies indeed indicate some difference. The difference in mass even reaches $\sim0.15M_{\odot}$, e.g. KIC 6064910, KIC 6766513, and KIC 10147635. This may caused by systematic errors associated with the grid and different inputs for $T_{\rm eff}$, [Fe/H]. Their estimations are based on coupling BeSPP to the GARSTEC grid. Their input $T_{\rm eff}$ are from SDSS recalibration and infra-red flux method calibration, both photometrically based, while ours are based on LAMOST spectroscopic observations, so that the above three stars have a significant $\sim300{\rm K}$ difference. In addition, they adopted a uniform [Fe/H] of $-0.2\pm0.3$ ${\rm dex}$, while we used values from LSP3.

\begin{figure}
\centering
     \includegraphics[width=13.0cm, angle=0]{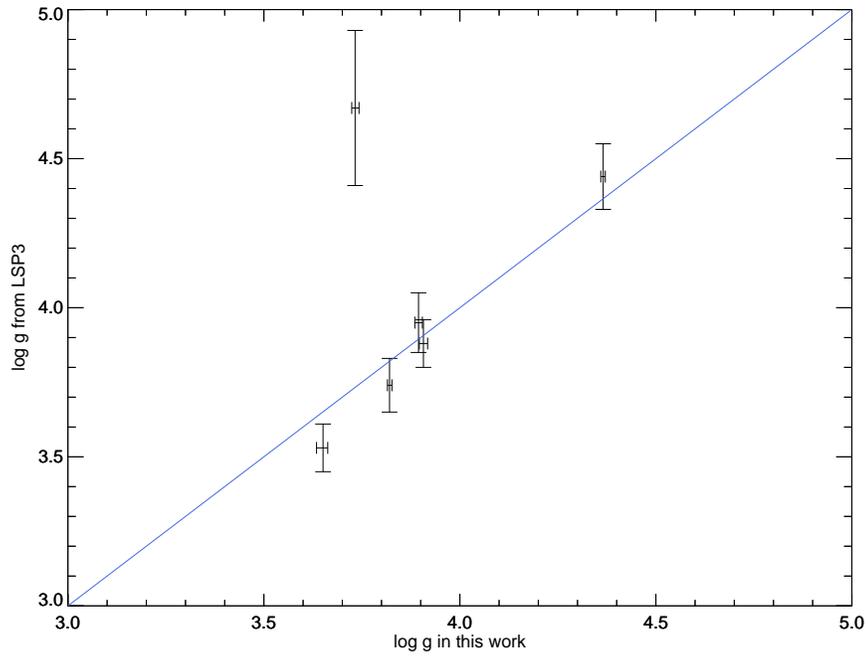}
  \caption{Comparison of $\log{g}$ from this work and LSP3. The blue solid line indicates equality.}
  \label{Fig:logg}
\end{figure}

Also, we compared the $\log{g}$ derived from LSP3 and this work. Figure \ref{Fig:logg} displays the comparison, where the blue solid line represents the equality. The results from the two work show a large difference, and the results from our work have small uncertainties, while those from LSP3 do not.

\section{Conclusions}
\label{sect:dac}
We performed data processing and asteroseismic analysis on six solar-like stars both observed by both the \emph{Kepler} mission and LAMOST, and combined stellar grid models to determine stellar fundamental parameters.

We derived asteroseismic parameters $\Delta\nu$ and $\nu_{\rm{max}}$ of the six targets using ACF and collapsed ACF. Individual mode frequencies $\nu_{nl}$ were extracted with Lorentzian profiles using a least-squares fit. For KIC 7107778 and KIC 10079226, we obtained the mode frequencies $\nu_{nl}$ with $l=0,1,2$. For KIC 6064910, KIC 6766513, and KIC 12069127, we obtained the mode frequencies $\nu_{nl}$ with $l=0,1$. Note that $\nu_{nl}$ with $l=1$ are more reliable than $\nu_{nl}$ with $l=0$ for these three stars, for the mix of $l=0$ and $l=2$ modes make them relatively ambiguous. Modes of KIC 10147635 were only obtained with $l=1$ due to relatively low S/N.

According to numerical solutions of the stellar models, we looked into the evolutionary stages of the six solar-like targets and categorized them into one main-sequence star (KIC 10079226) and five sub-giant stars (the others), four of which show strong characteristics of p and g mixed modes. Grid modeling indicates that the five sub-giant stars are in the range of $1.2\sim1.5$ $M_{\odot}$ and $2\sim3$ $R_{\odot}$, and the main-sequence star has corresponding values of $1.08M_{\odot}$ and $1.14R_{\odot}$. Their ages are in the range of $3\sim5$ $\rm{Gyr}$. Ages of most stars can reach an accuracy under 1 $\rm{Gyr}$, reflecting the capabilities of asteroseismology, but KIC 7107778 and KIC 10079226 have larger uncertainties in age.

In this work, we do not use individual oscillation frequencies to constrain stellar models. $\Delta\nu$ and $\nu_{\rm{max}}$ have shown their potential on constraining models, but individual oscillation frequencies can constrain them in a more powerful way. This will be presented in our follow-up research.

\normalem
\begin{acknowledgements}
The author would like to thank the whole \emph{Kepler} and LAMOST team, who make these results possible. We also greatly acknowledge the support provided by National College Students Innovation Training Program. This work is supported by grants 11273007 and 10933002 from the National Natural Science Foundation of China, the Joint Research Fund in Astronomy (U1631236) under cooperative agreement between the National Natural Science Foundation of China (NSFC) and Chinese Academy of Sciences (CAS), and the Fundamental Research Funds for the Central Universities.
\end{acknowledgements}

\bibliographystyle{raa}
\bibliography{bibtex}

\clearpage

\begin{table}
\bc
\begin{minipage}[]{200mm}
\caption[]{Observations of six solar-like targets.}\label{tab1}\end{minipage}
\setlength{\tabcolsep}{6pt}
\small
 \begin{tabular}{ccccccc}
  \hline\hline\noalign{\smallskip}
NO. & KIC& \emph{Kepler} data     & $T_{\rm{eff}}$&$\log{g}$  &$\rm{[Fe/H]}$&\\
  &   & (Q)&$(\rm{K})$     &$(\rm{dex})$&$(\rm{dex})$ &\\
  \hline\noalign{\smallskip}
1 & 6064910  & 7.1$\sim$11.3 & 6286$\pm$59 &3.74$\pm$0.09 &-0.33$\pm$0.08\\ 
2 & 6766513  & 7.1$\sim$11.3 & 6294$\pm$61 &3.88$\pm$0.08 &-0.18$\pm$0.07\\ 
3 & 7107778  & 7.1$\sim$11.3 & 5118$\pm$191  &3.53$\pm$0.08 &0.07$\pm$0.13 \\ 
4 & 10079226 & 7.1$\sim$10.3 & 5889$\pm$61  &4.44$\pm$0.11 &0.11$\pm$0.06 \\ 
5 & 10147635 & 7.1$\sim$11.3 & 5814$\pm$59 &4.67$\pm$0.26 &-0.08$\pm$0.18 \\ 
6 & 12069127 & 7.1$\sim$11.3 & 6305$\pm$64 &3.95$\pm$0.10 &0.16$\pm$0.08 \\ 
  \noalign{\smallskip}\hline
\end{tabular}
\ec
  \tablecomments{0.86\textwidth}{``Q'' represents three-month-long observation quarter. Atmospheric parameters were derived by LSP3 \citep{2015yCat..74480822X}.}
\end{table}

\begin{table}
\bc
\begin{minipage}[]{200mm}
\caption[]{$\Delta\nu$ and $\nu_{\rm{max}}$ of six solar-like targets.}\label{tab:dnu+numax}\end{minipage}
\setlength{\tabcolsep}{6pt}
\small
 \begin{tabular}{ccccccccccccc}
  \hline\hline\noalign{\smallskip}
  &KIC & $\Delta\nu^{(a)}$   &$\Delta\nu^{(b)}$  & $\nu_{\rm{max}}^{(a)}$ & $\nu_{\rm{max}}^{(b)}$\\
  &    & $(\mu\rm{Hz})$               & $(\mu\rm{Hz})$                  &$(\mu\rm{Hz})$              & $(\mu\rm{Hz})$\\
  \hline\noalign{\smallskip}
1 & 6064910   &43.90$\pm$0.40  &43.90$\pm$0.80  &733$\pm$37   &721$\pm$43\\
2 & 6766513   &51.30$\pm$0.99  &51.30$\pm$1.10  &832$\pm$90   &883$\pm$84\\
3 & 7107778  &31.40$\pm$0.34  &31.40$\pm$0.34  &540$\pm$19   &540$\pm$19\\
4 & 10079226  &116.40$\pm$0.86 &116.4$\pm$1.9   &2588$\pm$135 &2689$\pm$93\\
5 & 10147635  &37.40$\pm$0.54  &37.40$\pm$0.50  &582$\pm$28   &634$\pm$20\\
6 & 12069127  &48.20$\pm$0.80  &48.20$\pm$0.90  &817$\pm$50   &829$\pm$41\\
  \noalign{\smallskip}\hline
\end{tabular}
\ec
\tablecomments{0.86\textwidth}{(a): results in this work; (b): results from \cite{2014ApJS..210....1C}.}
\end{table}

\begin{table}
\bc
\begin{minipage}[]{200mm}
\caption[]{Oscillation frequencies of the six stars.}\label{frequency}\end{minipage}
\setlength{\tabcolsep}{5pt}
\small
 \begin{tabular}{ccccccccccccc}
\hline	\hline																									
	&	KIC 6064910			&	KIC 6766513			&	KIC 7107778			&	KIC 10079226			&	KIC 10147635			&	KIC 12069127				\\
\hline																										
$l=0$	&$	665.58	\pm	0.59	$&$	768.62	\pm	0.28	$&$	460.31	\pm	0.67	$&$	2024.15	\pm	0.31	$&	...			&$	587.64	\pm	1.22	$	\\
	&$	709.79	\pm	0.53	$&$	820.49	\pm	0.61	$&$	491.94	\pm	0.13	$&$	2137.90	\pm	0.33	$&				&$	633.97	\pm	0.84	$	\\
	&$	753.53	\pm	0.50	$&$	871.94	\pm	0.31	$&$	523.40	\pm	0.09	$&$	2252.89	\pm	0.57	$&				&$	680.88	\pm	0.28	$	\\
	&$	798.55	\pm	0.48	$&$	923.14	\pm	0.58	$&$	555.42	\pm	0.10	$&$	2367.81	\pm	1.06	$&				&$	728.91	\pm	0.45	$	\\
	&$	841.26	\pm	0.46	$&$	974.26	\pm	0.21	$&$	587.35	\pm	0.29	$&$	2484.26	\pm	0.22	$&				&$	778.03	\pm	0.49	$	\\
	&	...			&$	1025.95	\pm	0.48	$&	...			&$	2601.37	\pm	1.38	$&				&$	827.56	\pm	0.51	$	\\
	&				&	...			&				&$	2717.09	\pm	0.66	$&				&$	874.58	\pm	0.99	$	\\
	&				&				&				&$	2833.06	\pm	0.30	$&				&$	922.56	\pm	0.48	$	\\
	&				&				&				&$	2949.83	\pm	0.67	$&				&$	969.42	\pm	0.93	$	\\
	&				&				&				&	...			&				&$	1019.04	\pm	0.18	$	\\
	&				&				&				&				&				&$	1066.98	\pm	1.79	$	\\
\hline																										
$l=1$	&$	545.73	\pm	0.40	$&$	649.63	\pm	0.58	$&$	477.10	\pm	0.09	$&$	1958.06	\pm	0.47	$&$	519.61	\pm	0.14	$&$	608.00	\pm	0.26	$	\\
	&$	608.76	\pm	0.37	$&$	695.94	\pm	0.35	$&$	507.93	\pm	0.07	$&$	2077.07	\pm	0.32	$&$	544.65	\pm	0.41	$&$	653.49	\pm	0.63	$	\\
	&$	642.22	\pm	0.16	$&$	743.03	\pm	0.58	$&$	539.75	\pm	0.07	$&$	2192.30	\pm	0.21	$&$	592.95	\pm	0.24	$&$	700.66	\pm	0.76	$	\\
	&$	677.58	\pm	0.41	$&$	785.60	\pm	0.17	$&$	570.05	\pm	0.03	$&$	2306.51	\pm	0.74	$&$	624.67	\pm	0.31	$&$	749.97	\pm	0.46	$	\\
	&$	735.46	\pm	1.11	$&$	808.18	\pm	0.30	$&	...			&$	2422.61	\pm	0.50	$&$	658.26	\pm	0.34	$&$	799.20	\pm	0.47	$	\\
	&$	776.81	\pm	0.61	$&$	847.75	\pm	0.61	$&				&$	2539.49	\pm	0.37	$&$	704.02	\pm	0.28	$&$	848.06	\pm	0.65	$	\\
	&$	817.95	\pm	0.14	$&$	895.52	\pm	1.02	$&				&$	2655.42	\pm	0.36	$&$	738.06	\pm	0.36	$&$	895.97	\pm	0.54	$	\\
	&$	863.69	\pm	0.43	$&$	948.01	\pm	0.12	$&				&$	2771.85	\pm	0.69	$&$	773.20	\pm	0.59	$&$	943.72	\pm	0.57	$	\\
	&$	906.02	\pm	0.13	$&$	997.60	\pm	0.41	$&				&$	2886.92	\pm	0.84	$&	...			&$	992.19	\pm	0.37	$	\\
	&$	948.37	\pm	2.01	$&	...			&				&$	3002.24	\pm	0.92	$&				&$	1041.39	\pm	0.75	$	\\
	&	...			&				&				&	...			&				&$	1090.01	\pm	0.64	$	\\
\hline																										
$l=2$	&	...			&	...			&$	457.10	\pm	0.54	$&$	2125.57	\pm	0.18	$&	...			&	...				\\
	&				&				&$	488.86	\pm	0.15	$&$	2243.93	\pm	0.61	$&				&					\\
	&				&				&$	520.78	\pm	0.07	$&$	2358.83	\pm	0.36	$&				&					\\
	&				&				&$	553.11	\pm	0.11	$&$	2476.61	\pm	0.82	$&				&					\\
	&				&				&$	584.18	\pm	0.14	$&$	2591.81	\pm	0.27	$&				&					\\
	&				&				&	...			&$	2707.97	\pm	1.39	$&				&					\\
	&				&				&				&$	2825.57	\pm	0.75	$&				&					\\
	&				&				&				&$	2939.12	\pm	0.65	$&				&					\\

  \noalign{\smallskip}\hline
\end{tabular}
\ec
\end{table}

\begin{table}
\bc
\begin{minipage}[]{200mm}
\caption[]{Input parameters of grid modelling.}\label{tab3}\end{minipage}
\setlength{\tabcolsep}{2.5pt}
\small
 \begin{tabular}{ccccccccccccc}
 \hline\hline\noalign{\smallskip}
  Variable & Value\\
  \hline\noalign{\smallskip}
  Mass$(M_\odot)$          & $0.8\sim1.8$, $\delta=0.02$\\
  $\rm{[Fe/H]}\rm{(dex)}$  & $-0.3\sim0.4$, $\delta=0.1$\\
  $\alpha_{\rm{MLT}}$ & 1.75, 1.84, 1.95\\
  \noalign{\smallskip}\hline
\end{tabular}
\ec
\end{table}

\clearpage

\begin{sidewaystable}[h]
\bc
\begin{minipage}[]{300mm}
\bigskip\bigskip\bigskip\bigskip\bigskip\bigskip\bigskip\bigskip\bigskip\bigskip\bigskip\bigskip\caption{Stellar parameters estimated through grid modeling.}\label{tab4}\end{minipage}
\setlength{\tabcolsep}{2pt}
\centering
 \begin{tabular}{lccccccccccccccc}
		\hline\hline
KIC & $M^{(a)}$     & $M^{(b)}$     & $M^{(c)}$     & $R^{(a)}$       & $R^{(b)}$       & $R^{(c)}$       & $t^{(a)}$ & $t^{(b)}$ & $t^{(c)}$ & $\log{g}^{(a)}$ & $\log{g}^{(b)}$ & $\log{g}^{(c)}$ & $L^{(a)}$          & $T_{\rm eff}^{(a)}$  \\
    & ($M_{\odot}$) & ($M_{\odot}$) & ($M_{\odot}$) & ($R_{\odot}$)   & ($R_{\odot}$)   & ($R_{\odot}$)   & (Gyr)     & (Gyr)     & (Gyr)     & (dex)           & (dex)           & (dex)           &($L_{\odot}$)       & (K)            \\
		\hline
6064910& $1.26^{+0.04}_{-0.02}$ & $1.40^{+0.15}_{-0.13}$ & $1.48^{+0.13}_{-0.12}$ & $2.29^{+0.03}_{-0.03}$ & $2.37^{+0.10}_{-0.09}$ & $2.36^{+0.10}_{-0.08}$ & $3.56^{+0.24}_{-0.22}$ & $3.70^{+0.80}_{-0.90}$ & $3.00^{+0.60}_{-0.80}$ & $3.82^{+0.01}_{-0.01}$ & $3.84^{+0.02}_{-0.02}$ & $3.86^{+0.02}_{-0.02}$ & $7.46^{+0.22}_{-0.38}$ & $6288^{+  64}_{-  57}$ \\
6766513& $1.26^{+0.04}_{-0.04}$ & $1.41^{+0.16}_{-0.15}$ & $1.35^{+0.16}_{-0.12}$ & $2.05^{+0.04}_{-0.04}$ & $2.13^{+0.09}_{-0.09}$ & $2.11^{+0.08}_{-0.08}$ & $3.94^{+0.52}_{-0.30}$ & $3.30^{+1.10}_{-1.00}$ & $4.00^{+0.90}_{-1.20}$ & $3.91^{+0.01}_{-0.01}$ & $3.93^{+0.02}_{-0.02}$ & $3.92^{+0.02}_{-0.02}$ & $5.99^{+0.39}_{-0.38}$ & $6301^{+  57}_{-  64}$ \\
7107778& $1.48^{+0.14}_{-0.14}$ & $1.39^{+0.17}_{-0.14}$ & $1.35^{+0.17}_{-0.14}$ & $3.05^{+0.09}_{-0.14}$ & $2.96^{+0.14}_{-0.13}$ & $2.92^{+0.14}_{-0.12}$ & $3.08^{+1.92}_{-0.92}$ & $4.10^{+1.20}_{-1.20}$ & $4.60^{+1.20}_{-1.30}$ & $3.65^{+0.01}_{-0.02}$ & $3.64^{+0.01}_{-0.01}$ & $3.64^{+0.01}_{-0.01}$ & $5.93^{+1.10}_{-0.97}$ & $5148^{+ 172}_{- 147}$ \\
10079226& $1.08^{+0.04}_{-0.02}$ & $1.15^{+0.09}_{-0.11}$ & $1.13^{+0.10}_{-0.11}$ & $1.14^{+0.01}_{-0.01}$ & $1.15^{+0.04}_{-0.04}$ & $1.15^{+0.04}_{-0.04}$ & $4.26^{+1.16}_{-1.00}$ & $2.40^{+2.10}_{-1.70}$ & $2.70^{+2.50}_{-1.90}$ & $4.37^{+0.01}_{-0.01}$ & $4.37^{+0.01}_{-0.01}$ & $4.37^{+0.01}_{-0.02}$ & $1.42^{+0.08}_{-0.08}$ & $5910^{+  51}_{-  83}$ \\
10147635& $1.34^{+0.08}_{-0.06}$ & $1.50^{+0.26}_{-0.15}$ & $1.52^{+0.25}_{-0.16}$ & $2.60^{+0.06}_{-0.06}$ & $2.71^{+0.14}_{-0.10}$ & $2.73^{+0.14}_{-0.11}$ & $3.34^{+0.54}_{-0.30}$ & $3.30^{+0.70}_{-1.30}$ & $3.20^{+0.70}_{-1.20}$ & $3.73^{+0.01}_{-0.01}$ & $3.75^{+0.02}_{-0.01}$ & $3.75^{+0.02}_{-0.02}$ & $6.95^{+0.45}_{-0.41}$ & $5814^{+  57}_{-  64}$ \\
12069127& $1.44^{+0.06}_{-0.04}$ & $1.38^{+0.14}_{-0.09}$ & $1.40^{+0.15}_{-0.09}$ & $2.25^{+0.04}_{-0.04}$ & $2.22^{+0.09}_{-0.07}$ & $2.22^{+0.09}_{-0.07}$ & $2.84^{+0.18}_{-0.32}$ & $3.70^{+0.70}_{-0.90}$ & $3.60^{+0.70}_{-1.10}$ & $3.89^{+0.01}_{-0.01}$ & $3.89^{+0.01}_{-0.01}$ & $3.89^{+0.02}_{-0.02}$ & $7.16^{+0.60}_{-0.26}$ & $6308^{+  83}_{-  51}$ \\
		\hline
\end{tabular}
\ec
\smallskip
\tablecomments{0.86\textwidth}{(a): results in this work; (b): results from \cite{2014ApJS..210....1C} with input $T_{\rm eff}$ from SDSS recalibration; (c): results from \cite{2014ApJS..210....1C} with input $T_{\rm eff}$ from Infra-Red Flux Method calibration.}
\end{sidewaystable}

\clearpage

\end{document}